\documentclass[twocolumn]{aastex701}
\accepted{to ApJ on 12/12/2025}
\begin{document}

%%%\title{X-ray Evolution of Young Stars: Rapid X-ray Dimming and Coronal Softening in Solar-Mass Stars Around 100 Myr}

\title{X-ray Evolution of Young Stars: Early Dimming and Coronal Softening in Solar-Mass Stars with Implications for Planetary Atmospheres}

\correspondingauthor{Konstantin Getman}
\email{kug1@psu.edu}

\author[0000-0002-6137-8280]{Konstantin V. Getman}
\affiliation{Department of Astronomy \& Astrophysics \\
Pennsylvania State University \\ 
525 Davey Laboratory \\
University Park, PA 16802, USA}
\email{kug1@psu.edu}

\author[0000-0002-5077-6734]{Eric D. Feigelson}
\affiliation{Department of Astronomy \& Astrophysics \\
Pennsylvania State University \\ 
525 Davey Laboratory \\
University Park, PA 16802, USA}
\affiliation{Center for Exoplanets and Habitable Worlds \\ 525 Davey Laboratory \\ The Pennsylvania State University \\ University Park, PA 16802, USA}
\email{e5f@psu.edu}

\author[0000-0003-4452-0588]{Vladimir S. Airapetian}
\affiliation{American University\\ 4400 Massachusetts Avenue NW\\ Washington, DC 20016 USA}
\affiliation{NASA/GSFC/SEEC\\ Greenbelt, MD, 20771, USA}
\email{vladimir.airapetian-1@nasa.gov}

\author[0000-0002-7371-5416]{Gordon P. Garmire}
\affiliation{Huntingdon Institute for X-ray Astronomy\\
LLC, 10677 Franks Road\\
Huntingdon, PA 16652, USA}
\email{g2p3g4@gmail.com}

%% Use the \collaboration command to identify collaborations. This command
%% takes an optional argument that is either a number or the word "all"
%% which tells the compiler how many of the authors above the command to
%% show. For example "\collaboration[all]{(DELVE Collaboration)}" wil include
%% all the authors above this command.
%%
%% Mark off the abstract in the ``abstract'' environment. 
\begin{abstract}
X-ray and ultraviolet (XUV) emission from young stars plays a critical role in shaping the evolution of planetary atmospheres and the conditions for habitability. To assess the long-term impact of high-energy stellar radiation, it is essential to empirically trace how X-ray luminosities and spectral hardness evolve during the first $\lesssim 1$~Gyr, when atmospheric loss and chemical processing are most active. This study extends the X-ray activity-mass-age analysis of $<$25~Myr stars by \citet{Getman2022} to ages up to $\sim 750$~Myr, using Gaia-based cluster memberships, new {\it Chandra} observations of five rich open clusters ($\sim 45$--100~Myr), and archival {\it ROSAT} and {\it Chandra} data for three older clusters ($\sim 220$--750~Myr). We find a mass-dependent decay in X-ray luminosity: solar-mass stars undergo a far more rapid and sustained decline, accompanied by coronal softening and the disappearance of hot plasma by $\sim 100$~Myr, compared to their lower-mass siblings. These trends in solar-mass stars are likely linked to reduced magnetic dynamo efficiency and diminished ability to sustain large-scale, high-temperature coronal structures. The trends are significantly stronger than predicted by widely used XUV-rotation-age relations. The revised trends imply systematically lower rates of atmospheric mass loss and water photolysis, as well as altered ionization environments and chemical pathways relevant to the formation of prebiotic molecules, for planets in close orbits around solar analogs. These effects persist throughout at least the $\lesssim 750$~Myr interval probed in this study.
\end{abstract}

%% Keywords should appear after the \end{abstract} command. 
%% The AAS Journals now uses Unified Astronomy Thesaurus (UAT) concepts:
%% https://astrothesaurus.org
%% You will be asked to selected these concepts during the submission process
%% but this old "keyword" functionality is maintained in case authors want
%% to include these concepts in their preprints.
%%
%% You can use the \uat command to link your UAT concepts back its source.
%%\keywords{\uat{Galaxies}{573} --- \uat{Cosmology}{343} --- \uat{High Energy astrophysics}{739} --- \uat{Interstellar medium}{847} --- \uat{Stellar astronomy}{1583} --- \uat{Solar physics}{1476}}

%% From the front matter, we move on to the body of the paper.
%% Sections are demarcated by \section and \subsection, respectively.
%% Observe the use of the LaTeX \label
%% command after the \subsection to give a symbolic KEY to the
%% subsection for cross-referencing in a \ref command.
%% You can use LaTeX's \ref and \label commands to keep track of
%% cross-references to sections, equations, tables, and figures.
%% That way, if you change the order of any elements, LaTeX will
%% automatically renumber them.

\section{Introduction} \label{sec:introduction} 
X-rays and ultraviolet (XUV)\footnote{While both bands are important to planetary atmospheres, large homogeneous observational samples spanning a wide range of stellar ages exist only in the X-ray band.} emitted by young stars play a crucial role in shaping their circumstellar environments. In particular, X-rays are a key source of ionization in protoplanetary disks, enabling the magnetorotational instability (MRI) that drives accretion \citep{Turner2014,Bai2016,Waggoner2022,Washinoue2024}, and shaping the turbulent conditions that can either hinder or promote planetesimal formation via processes such as the streaming instability \citep{Johansen2011,XuBai2022,Lim2024}. Stellar X-rays also initiate nonequilibrium gas-phase ion-molecule chemistry, influencing the abundances of molecular ions such as HCO$^+$ and N$_2$H$^{+}$, and contributing to the production of observed [NeII] emission lines from disk surfaces \citep{Glassgold2000,Bergin2007,Cleeves2015,Huang2015,Cleeves2017,Espaillat2023}. High-energy particles from powerful X-ray flares, at least in the solar nebula, produce spallogenic radionuclides in disk solids \citep{McKeegan00,Yang2024}. XUV radiation is the principal cause of photoevaporation of disks after a few to several million years of irradiation \citep{Gorti2009,Owen2012,Alexander2014,Picogna2021,Ercolano2022, Sellek2022,Ercolano2023,  Manara2023, Cecil2024}.

High-energy irradiation from stellar X-rays and extreme ultraviolet (EUV) photons is widely considered the dominant driver of atmospheric mass loss from close-in young exoplanets \citep{Airapetian2017,Garcia-Sage2017,Owen2019}. More recently, energetic particles from stellar winds and flares  --- though observationally less constrained --- have also been proposed to contribute significantly to atmospheric escape \citep{Airapetian2020,  Harbach2021,Alvarado-Gomez2022, Hazra2022,Brunn2024}. The observed structure of the so-called ``Neptune desert'', and potentially the $\sim$1.8~R$_\oplus$ radius valley \citep{Mazeh2016,Fulton2017}, may reflect the time-dependent evolution of high-energy stellar output, particularly in the form of X-ray and EUV emission during the early stages of planetary system evolution \citep{Tu2015, Johnstone2021}.

High-energy-driven atmospheric escape may lead to abiotic oxygen buildup. Photodissociation of water vapor in young, irradiated planets can result in hydrogen loss to space and potential accumulation of residual O$_2$ \citep{Luger2015,Wordsworth2014}. If surface sinks are inefficient or depleted, this O$_2$ may reach levels resembling biological oxygenation but with surface conditions hostile to life, possibly suppressing prebiotic chemistry. However, strong hydrodynamic outflows may entrain and remove oxygen, preventing its accumulation \citep{Johnstone2020}. These processes are now explored with advanced models of atmospheric escape and photochemistry \citep{Kubyshkina2021, Ketzer2022, Wordsworth2022}.

In addition to atmospheric escape and potential oxygen buildup, early high-energy radiation from host stars may also sterilize or chemically alter planetary surfaces, initiate prebiotic chemical pathways in exoplanetary atmospheres, and influence the early evolution and detectability of young planetary atmospheres \citep[e.g.,][]{Airapetian2016, Rimmer2018, Dong2017, Rimmer2023,Kobayashi2023}.

Understanding the long-term impact of high-energy stellar radiation on disks and planets requires empirical characterization of how stellar X-ray luminosity and spectral hardness evolve with time --- from the earliest ($<1$--$5$~Myr) phases of planet formation in gaseous disks, to the later ($>5$~Myr) epochs when primordial and secondary planetary atmospheres are most vulnerable to alteration and escape. Such studies must span a broad range of stellar ages; for example, focusing on a single nearby region like Taurus or a single star cluster like the Pleiades is insufficient. Open clusters in the $10$--$1000$~Myr range, though often more distant and less studied, are crucial. In addition, X-ray samples should include stars with a range of masses and rotation rates to probe the diversity of radiation environments that may have shaped older planetary systems.

Several early studies have examined the mass-dependent evolution of X-ray luminosity in young stars \citep[e.g.,][]{Gudel2004,PreibischFeigelson2005,Pillitteri2006,Nunez2016}. These studies considered nearby stellar cluster samples observed with various X-ray telescopes (ROSAT, XMM, and {\it Chandra}), using heterogeneous methodologies and relying on cluster membership information that predates the Gaia era. Cluster memberships based on Gaia distances and proper motions are critical for assessing the influence of undetected members on the derived X-ray luminosity distributions.  

Recently, \citet{Getman2022} conducted a more homogeneous study of rich stellar clusters aged $<1$--25 Myr, leveraging joint {\it Chandra} and Gaia data to include both X-ray-emitting and non-emitting members.  They found that, within each stellar mass stratum, X-ray emission remains nearly constant during the first few Myr, likely due to the presence of extended coronae that are relatively insensitive to the size changes of contracting pre-main sequence stars. Over the subsequent [7--25]~Myr period, however, the emission declines: $L_X \propto t^{-0.6}$ for $<1$~M$_{\odot}$ stars due to evolving convection-driven dynamos and coronal extents; $L_X \propto t^{-1.8}$ for [1--2]~M$_{\odot}$ stars due to the development of solar-type tachocline dynamos; and $L_X \propto t^{-4}$ or steeper for more massive stars that often lack magnetic dynamos in their fully radiative interiors. A limitation of this study is the restriction of the $L_X$--$t$ analysis to stars more massive than $0.75$~M$_{\odot}$, set by the mass completeness threshold of their {\it Chandra}-Gaia stellar samples.

\citet{Johnstone2021} developed a framework combining a modified stellar angular momentum evolution model \citep{Gallet2015} --- including star-disk coupling, wind-driven spin-down, and internal angular-momentum transport --- with empirical relations connecting rotation (via Rossby number) to fractional X-ray luminosity ($R_X = L_X/L_{bol}$). By calibrating this model against rotational and X-ray data from clusters spanning a wide age range, they distinguish between saturated and non-saturated dynamo regimes. This work provides predicted $L_X$ evolution for different stellar masses and rotation histories, offering a complementary approach to observational studies.

Our current work extends the X-ray activity-mass-age analysis of $<$25~Myr stars by \citet{Getman2022} to ages up to $\sim 750$~Myr, using Gaia-based cluster memberships, new {\it Chandra} observations of five rich open clusters ($\sim 45$--100~Myr), and archival ROSAT and {\it Chandra} data for three older clusters ($\sim 220$--750~Myr). We perform joint X-ray-Gaia analyses of cluster distances, ages, stellar memberships, and stellar properties. We evaluate mass-stratified relationships between $L_X$ and age, as well as coronal plasma temperature and age; interpret the underlying mechanisms; compare our $L_X$-$t$ results with the semi-empirical study of \citet{Johnstone2021}; and discuss the potential implications for high-energy ionizing radiation effects on young planetary environments. 

The paper is organized as follows. Sections \ref{sec:targets_chandra}--\ref{sec:Xray_luminosities} describe the sample of five $\sim 45$--100~Myr clusters recently observed with {\it Chandra}, including X-ray data reduction, cluster membership determination, stellar and cluster characterization, and the derivation of stellar X-ray luminosities and upper limits. Section \ref{sec:3more_older_clusters} discusses the addition of three older clusters ($\sim 220$--750~Myr) previously observed with ROSAT or {\it Chandra}, including identification and characterization of their members. Section \ref{sec:resuts} presents results on the temporal evolution of stellar X-ray luminosity and coronal plasma temperature. In Section \ref{sec:discussion}, we discuss the underlying physical mechanisms driving the observed X-ray evolution and consider potential implications for young planets. Supplementary tables and figures are provided in Appendix \ref{sec:appendix}.

\begin{deluxetable*}{lrrrrcrcc}
%\tablenum{2}
\tabletypesize{\footnotesize}
\tablecaption{Five Open Clusters with Ages $< 150$~Myr\label{tab:cluster_props}}
\tablewidth{0pt}
\tablehead{
\colhead{Region} & \colhead{$l$} &
\colhead{$b$} & \colhead{$N_{mem}$} &  \colhead{$D$} & \colhead{$A_{V}$} & \colhead{Age} & \colhead{$N_{>0.1M_{\odot}}$} & \colhead{$M_{lim}$}\\
\colhead{} & \colhead{(deg)} &  \colhead{(deg)} & & \colhead{(pc)} & \colhead{(mag)} & \colhead{(Myr)} & \colhead{} & \colhead{(M$_{\odot}$)}\\
\colhead{(1)} & \colhead{(2)} & \colhead{(3)} & \colhead{(4)} & \colhead{(5)} & \colhead{(6)} & \colhead{(7)} & \colhead{(8)} & \colhead{(9)}
}
%\decimalcolnumbers
\startdata
NGC 6242 & 345.4561 & 2.4667 & 760 (546) & $1280 \pm 15$ & 1.2 & 45 (30 - 65) & 3185 & $<0.75$\\
Trumpler 3 & 138.0154 & 4.5310 & 207 (105) & $680 \pm 2$ & 0.9 & 45 (35 - 60) &518 & $<0.75$\\
NGC 6204 & 338.5562 & -1.0335 & 468 (403) & $1205 \pm 16$ & 1.4 & 75 (60 - 120) & 1723 & $<0.75$\\
NGC 2353 & 224.6811 & 0.4066 & 229 (165) & $1206 \pm 10$ & 0.4 & 100 (75 - 130) & 883 & $\sim 0.75$\\
NGC 2301 & 212.5549 & 0.2830 & 478 (312) & $874 \pm 5$ & 0.2 & 100 (70 - 150) & 1370 & $<0.75$\\
\enddata
\tablecomments{Column 1: Cluster name. Columns 2-3: Galactic coordinates for the cluster center in degrees. Column 4: Total number of cluster members identified across the entire Chandra fields --- 2,142 X-ray and non-X-ray stars across the five clusters (presented later in Table \ref{tab:stellar_props_open_clusters}). The values in parentheses denote 1,531 members located in the central regions of the clusters that were included in the analyses of the X-ray-mass-age relations. Columns 5: Median cluster distance from the Sun based on Gaia data, with the associated 68\% bootstrap error. Column 6: Cluster average extinction in visual band. Column 7: Cluster age, where the best-fit value corresponds to the PARSEC PMS isochrone that most closely follows the empirical Gaia color-magnitude sequence as traced by a locally weighted regression curve; values in parentheses indicate approximate lower and upper age limits, based on visual identification of isochrones that begin to noticeably diverge from the data. Column 8: Initial mass function-based estimate of the intrinsic total number of cluster members down to 0.1~M$_{\odot}$ within the {\it Chandra} fields of view. Column 9: Mass completeness limits for our cluster member samples. The values in Columns 4-9 were obtained from the analyses described in Sections \S\S \ref{sec_membership} and \ref{sec:cluster_member_properties}.} 
\end{deluxetable*}

%% The "ht!" tells LaTeX to put the figure "here" first, at the "top" next
%% and to override the normal way of calculating a float position.
%% The asterisk after "figure" tells the compiler to span multiple columns
%% if a two column style is selected.

\begin{figure*}[ht!]
\plotone{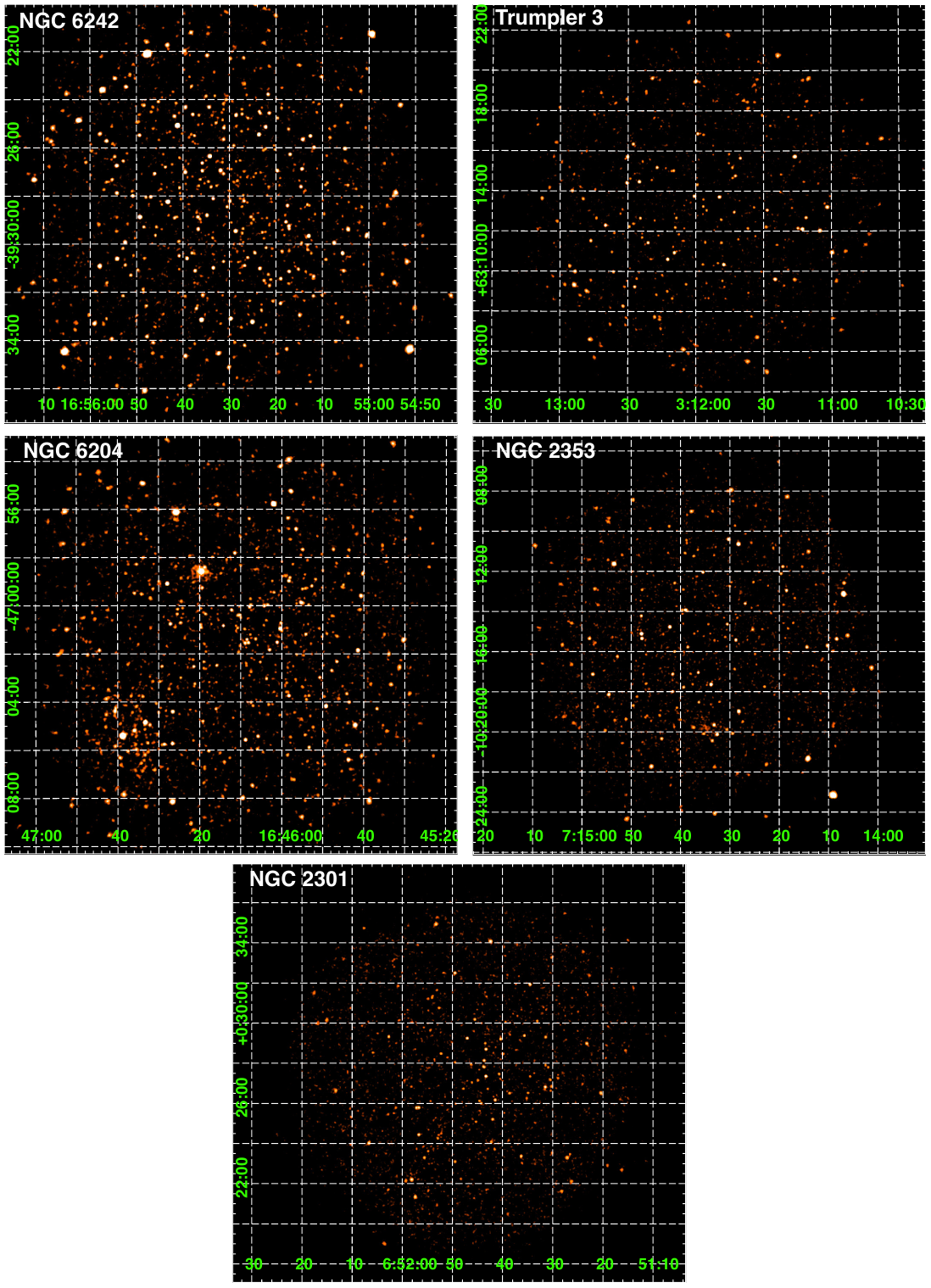}
\caption{Adaptively smoothed, low-resolution mosaic images of the five open clusters observed with {\it Chandra} ACIS-I, shown in the $(0.5-8)$~keV energy band, revealing a few thousand of the brighter X-ray sources across the fields.
\label{fig:chandra_images}}
\end{figure*}

\section{ {\it Chandra} Target Sample of Five $\sim 45$--100~Myr Clusters and Data Reduction} \label{sec:targets_chandra}

Our new sample consists of five nearby open clusters with ages ranging from 45 to 100~Myr, listed in Table~\ref{tab:cluster_props} in order of increasing age. These clusters were selected from the literature based on their age, proximity, compactness, and stellar richness. Table~\ref{tab:cluster_props} provides the number of {\it Chandra}--Gaia-based cluster members, distances, extinctions, and age estimates, as well as the approximate mass completeness limits of the member samples, as derived from the analyses presented in subsequent sections.

Table~\ref{tab:log_chandra_observations} in the Appendix summarizes the 63 {\it Chandra} observations of these clusters analyzed in this paper. The observations were conducted between 2021 and 2025 as part of a large GO/GTO program (PIs: Getman and Garmire). All observations employed the Advanced CCD Imaging Spectrometer imaging array \citep[ACIS-I;][]{Garmire03}. Each pointing used the $17\arcmin \times 17\arcmin$ field of view provided by the four contiguous CCD chips of ACIS-I, centered on the respective cluster cores. All data were acquired in Very Faint Timed Exposure mode.

The X-ray analysis methods used here are adapted from our previous investigations of young stellar populations \citep[e.g.,][]{Kuhn2013a, Getman17, Townsley2019, Getman2022}, and are summarized below. The Level 1 event lists generated by the Chandra X-ray Center pipeline were reprocessed and calibrated using standard CIAO tools (version 4.17) and calibration database CALDB 4.12.0. Source detection and characterization were performed using the {\it ACIS Extract} software package (version 5658 2022-01-25) and associated tools \citep{Broos10, Broos2012}, which offer high sensitivity and reliability in identifying faint X-ray sources within {\it Chandra} images. Candidate sources were initially identified as statistically significant enhancements in a smoothed representation of the merged X-ray field, generated via maximum likelihood reconstruction techniques that incorporate the spatially varying {\it Chandra} point-spread function (PSF). Background levels were estimated locally and refined iteratively for each candidate. This detection process was carried out on images co-aligned and merged from multiple {\it Chandra} observations.

To refine the absolute astrometry of each field, systematic offsets between bright X-ray detections and their Gaia DR3 stellar counterparts are removed, yielding sub-arcsecond positional accuracy for sources in the central parts of the detector. For each candidate source, X-ray events are extracted from regions tailored to the local PSF, typically encompassing 90\% of the expected encircled energy. Local background contributions are estimated and subtracted, producing net source counts. Exposure times are derived from the merged exposure map at each source position. A suite of source properties is then computed from the extracted photon lists, including net count rates corrected for effective exposure and PSF wings; positional uncertainties based on off-axis angle and source strength; statistical significance of detection using Poisson-based probabilities; temporal variability on short- and long-timescales assessed via the Kolmogorov-Smirnov test; apparent photometric fluxes; and the median photon energy of the net source events.

A total of 7,995 candidate X-ray sources were detected in the {\it Chandra} ACIS-I images of the five open clusters. A few thousand brighter sources are readily visible in the low-resolution, adaptively smoothed {\it Chandra} ACIS-I mosaics shown in Figure~\ref{fig:chandra_images}. Table~\ref{tab:xray_photometry} in the Appendix presents the basic X-ray properties of all detected sources, including their positions, net counts, apparent photometric fluxes, and the median energy of the net source photons. The catalog comprises young cluster members, extragalactic background sources, Galactic field stars, and a small number of likely spurious detections. The final column of the table indicates the source classification, as described in the following sections.

It is important to note that the south-eastern corner of the {\it Chandra} image of NGC~6204 contains a separate, more distant, and consequently more compact cluster, Hogg~22 (Figure~\ref{fig:chandra_images}). To avoid contamination, this entire south-eastern region is excluded from the subsequent membership analysis and stellar characterization of NGC~6204.

\begin{figure*}[ht!]
\plotone{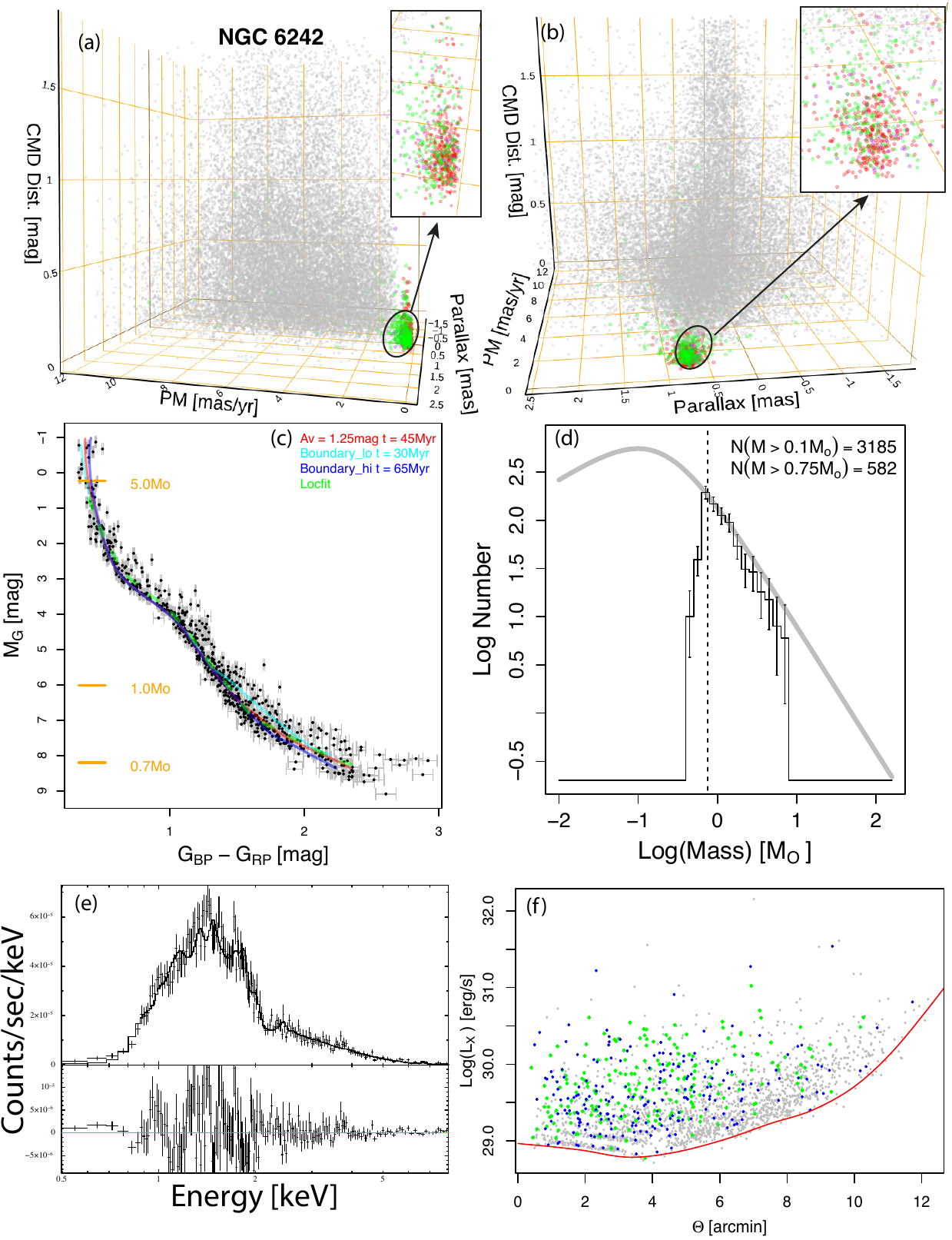}
\caption{Summary of member identification and properties for NGC 6242. (a,b) Final likely members: non-X-ray (Groups 1 and 2, red), X-ray (Groups 5 and 6, green), and additional members outside the {\it Chandra} field from \citet{Cantat-Gaudin2020} ( purple), compared with likely non-members (grey). Shown in 3D space defined by proper motion distance from the cluster's motion centroid, parallax, and photometric offset from the empirical CMD sequence. Insets show zoomed-in member views. (c) CMD for Groups 1, 2, 5, and 6 plus \citet{Cantat-Gaudin2020} members (black points with $G_{BP} - G_{RP}$ errors in grey). Overlaid are the best-fit PARSEC isochrone (red), bounding isochrones (cyan/blue), and the smoothed empirical sequence from {\it locfit} (dashed green). (d) IMF histogram for Groups 1, 2, 5, and 6 across the {\it Chandra} field (black line with 95\% Poisson error bars). Theoretical IMF from \citet{Maschberger2013} shown in grey. The dashed vertical line marks $M = 0.75$ M$_{\odot}$; the legend lists the observed $M > 0.75$ M$_{\odot}$ count and extrapolated population down to 0.1 M$_{\odot}$. (e) Stacked {\it Chandra} spectrum of Group 5 and 6 stars with best-fit two-temperature plasma model. (f) X-ray luminosity vs. angular distance from cluster center. Group 5 (green), Group 6 (blue), and likely contaminants (Groups 7–10, grey) are shown. A red local regression to the lower envelope is used to estimate X-ray upper limits for undetected (Groups 1 and 2) cluster members. Analogous figures for other clusters appear in the Appendix (Figures~\ref{fig:mem_TRUMPLER3}-\ref{fig:mem_NGC2301}).
\label{fig:mem_NGC6242}}
\end{figure*}

\section{Identifying Members of Five $\sim 45$--100~Myr Clusters} 
\label{sec_membership}

The identification of stellar members in the five open clusters studied here builds on the methodology developed in our previous work \citep{Getman2022}, adapted to this new dataset. Membership is established by jointly analyzing the X-ray detections cataloged in Table~\ref{tab:xray_photometry} (in the Appendix section) and stellar data from the {\it Gaia} Data Release 3 (DR3) \citep{GaiaMission2016, GaiaEDR32021}. DR3 provides high-precision astrometric and photometric measurements across the entire sky in the optical band. While all confirmed members possess Gaia counterparts, only a subset are detected in X-rays, so the final membership lists include both X-ray and non-X-ray stars located within the footprints of the {\it Chandra} observations.

The member selection procedure is summarized below, with an example shown for NGC 6242 in Figure~\ref{fig:mem_NGC6242}, and similar results for the remaining clusters presented in the Appendix (Figures~\ref{fig:mem_TRUMPLER3}-\ref{fig:mem_NGC2301}).

We begin by matching sources from the \textit{Chandra} X-ray catalog to Gaia~DR3 using a fixed cross-matching radius of 1\arcsec. This radius, consistent with our previous Gaia--\textit{Chandra} studies of stellar clusters, adequately accounts for \textit{Chandra}/ACIS-I positional uncertainties, which typically range from 0.1\arcsec\ in the central field to $\gtrsim$1\arcsec\ near the edges \citep{Kuhn19,Getman2019,Kuhn2022}. To estimate the rate of spurious matches, we shifted each \textit{Chandra} catalog by 1\arcmin\ north and repeated the matching. The resulting false-match fractions, 7--26\%, represent the expected maximum contamination among the original ACIS--Gaia pairs. However, 93--100\% of these spurious matches are linked to Gaia sources unrelated to the clusters, implying an extremely low probability of falsely identifying nonmembers as X-ray--Gaia cluster members. 

Across the five clusters, the median angular separation for true matches (0.3--0.5\arcsec) is notably smaller than that obtained for the shifted catalogs (0.7--0.9\arcsec), indicating that when both genuine and chance counterparts fall within the adopted matching radius, the genuine counterpart is more likely to be selected. In a few cases --- typically for X-ray sources at the outermost edges of the \textit{Chandra} field with positional errors $>1$--$2$\arcsec --- a genuine Gaia counterpart may be missed. Such rare misclassifications do not affect our results, which, as described below, are based solely on confirmed cluster members located within the central ACIS regions used for the mass-stratified X-ray activity analysis.

The matched X-ray/optical sample forms a core subset for defining cluster kinematics.

Since young, low-mass stars exhibit elevated X-ray activity --- often 100–1000 times that of older main-sequence counterparts \citep[e.g.,][their Figure 7]{Pillitteri2006} --- these X-ray-detected stars with Gaia counterparts provide a reliable starting point for identifying likely cluster members. We select stars from this group that have precise proper motion measurements (with errors in right ascension proper motion, $\sigma_{\mu_\alpha}$, less than 0.2 mas/yr). These are placed in proper motion space, and we define a circular selection region that encompasses the densest concentration of stars. The radius of this proper motion circle, $R_{cl,pm}$, is set to 1.3 times the normalized median absolute deviation (MADN), typically enclosing $60-90$\% of the Gaia+X-ray sample. Although this conservative radius may exclude some genuine members with poorer astrometry, it minimizes contamination from unrelated field stars.

Using this cleaned sample, we estimate the cluster distance from Gaia parallaxes. Following \citet{Getman2022}, we adopt the median parallax of the selected stars and compute the distance as its inverse. Since all stars have positive parallaxes, the results are consistent whether one takes the median parallax or the median of individual distances. Distance uncertainties are derived through bootstrap resampling of the parallax distribution and are listed in Table~\ref{tab:cluster_props}. Probabilistic distance estimates from \citet{Bailer-Jones2021} are systematically lower by $\sim 2$\%, primarily because, as in \citet{Getman2022}, we do not apply a global zero-point parallax correction. This small offset has negligible impact on the derived cluster ages and stellar X-ray luminosity functions.

To supplement the parallax-based distance estimation, we also define a parallax spread that serves as a membership boundary in parallax space. Specifically, we calculate this spread as twice the normalized median absolute deviation of the parallaxes, i.e., $PlxSpread = 2 \times MADN$. This range provides a robust criterion for selecting additional members whose parallaxes are consistent with the cluster distance. Combining this parallax spread with the proper motion constraints described earlier, we classify all Gaia sources --- both those with and without X-ray counterparts --- located within the {\it Chandra} fields of view into the following ten groups (see Table~\ref{tab:summary_source_classes} for a concise description of these classes).

Groups 1, 2, and 3 consist of Gaia sources without X-ray detections. They are distinguished based on the precision of their proper motion measurements. Group 1: Gaia-only sources with high-precision proper motions ($\sigma_{\mu_{\alpha}} < 0.2$ mas/yr). Group 2: Sources with moderate precision ($0.2 < \sigma_{\mu_{\alpha}} < 0.4$ mas/yr). Group 3: Sources with low-precision proper motions ($\sigma_{\mu_{\alpha}} > 0.4$ mas/yr). All of these stars have proper motion vectors falling within the defined cluster radius in proper motion space ($R_{cl,pm}$) and have parallax confidence intervals that intersect the PlxSpread band centered on the cluster median parallax. Upon further inspection of their photometric properties, we find that Group 1 primarily includes higher mass cluster members (typically $0.7-8$~M$_{\odot}$), while Group 2 contains a mixture of lower-mass members ($0.3-0.7$~M$_{\odot}$). Group 3 is more ambiguous: it likely includes both faint, very low-mass cluster members and a higher fraction of field contaminants, both foreground and background.

Group 4 includes the remainder of Gaia sources within the {\it Chandra} fields that are not detected in X-rays and do not meet the proper motion and parallax criteria for cluster membership. These are presumed to be unrelated foreground or background field stars.

Groups 5 and 6 consist of X-ray sources with Gaia counterparts whose parallaxes fall within the PlxSpread interval, indicating distances consistent with cluster membership. These two groups are distinguished based on proper motion. Group 5 --- sources whose proper motions lie within the cluster proper motion defined radius ($R_{cl,pm}$). Group 6 --- sources with proper motions outside the $R_{cl,pm}$ region. Photometric properties from Gaia suggest that Group 5 primarily contains likely members with stellar masses in the range of $0.5-2$ solar masses. Group 6 likely includes a mix of similarly massive members with somewhat discrepant kinematics and a modest population of contaminant stars.

Groups 7 and 8 are defined by having parallaxes that lie significantly outside the cluster parallax range. Group 7 --- X-ray+Gaia sources whose lower parallax error bars lie above the PlxSpread range, suggesting foreground objects. Group 8 --- X-ray+Gaia sources whose upper parallax error bars fall below the PlxSpread range, indicating probable background stars. These two groups predominantly consist of field stars not physically associated with the clusters.

Groups 9 and 10 account for sources with incomplete Gaia information. Group 9: X-ray sources with Gaia matches but lacking usable astrometric measurements (e.g., missing or unreliable proper motions and/or parallaxes). Some of these may be very low-mass members, while others are likely contaminants. Group 10: X-ray sources without any Gaia counterpart. These are overwhelmingly dominated by background extragalactic sources (e.g., AGN) or false detections due to noise fluctuations in the X-ray images.

\begin{deluxetable*}{ccccc}
\tabletypesize{\footnotesize}
\tablecaption{Summary of Source Classifications\label{tab:summary_source_classes}}
\tablewidth{0pt}
\tablehead{
\colhead{Group} & \colhead{X-ray} &
\colhead{Comment 1} & \colhead{Comment 2} &  \colhead{$N$}\\
\colhead{(1)} & \colhead{(2)} & \colhead{(3)} & \colhead{(4)} & \colhead{(5)}
}
%\decimalcolnumbers
\startdata
1  &    No  &    $\sigma_{\mu_{\alpha}} < 0.2$ mas/yr;  inside $R_{cl,pm}$; intersects PlxSpread; young CMD locus & Members $>0.7$ M$_{\odot}$ & 967 (551)\\
2  &    No  &    $0.2 < \sigma_{\mu_{\alpha}} < 0.4$ mas/yr;  inside $R_{cl,pm}$; intersects PlxSpread; ; young CMD locus & Members $<0.7$ M$_{\odot}$ & 225 (110)\\
3  &    No  &    $\sigma_{\mu_{\alpha}} > 0.4$ mas/yr;  inside $R_{cl,pm}$; intersects PlxSpread & Contaminants or Low-mass Members & 290\\
4  &    No  &    none of the above & Contaminants & 87791\\
5  &    Yes &    any $\sigma_{\mu_{\alpha}}$;  inside $R_{cl,pm}$; intersects PlxSpread; young CMD locus & Members $0.5-2$ M$_{\odot}$ & 526 (483)\\
6  &    Yes &    any $\sigma_{\mu_{\alpha}}$;  outside $R_{cl,pm}$; intersects PlxSpread; young CMD locus & Members $0.5-2$ M$_{\odot}$ & 424 (387)\\
7  &    Yes &    any $\sigma_{\mu_{\alpha}}$;  any $R_{cl,pm}$; above PlxSpread & Foreground Stars & 113\\
8  &    Yes &    any $\sigma_{\mu_{\alpha}}$;  any $R_{cl,pm}$; below PlxSpread & Background Stars & 325\\
9  &    Yes &    lacks Gaia astrometry & Mainly Contaminants & 184\\
10 &    Yes &    lacks Gaia counterpart & AGN or False Detection & 5759\\
\enddata
\tablecomments{Column 1: Source class. Column 2: Indicates whether sources were detected by Chandra. Column 3: Brief definition of the source class. Column 4: Cluster membership summary. Column 5: Total number of sources (summed across the five clusters) corresponding to this class. For Groups 1, 2, 5, and 6, all the numbers correspond to the additional correction for the location on the Gaia color-magnitude diagram. The numbers in parentheses indicate stars located within the central regions of the clusters ($R < 7.5\arcmin$), employed in the activity-mass-age science analyses.}
\end{deluxetable*}

\begin{deluxetable*}{ccccccccccc}
%\tablenum{2}
\tabletypesize{\footnotesize}
\tablecaption{Stellar Properties of X-ray and Non-X-ray Members of Five $\sim 45$--100~Myr Open Clusters\label{tab:stellar_props_open_clusters}}
\tablewidth{0pt}
\tablehead{
\colhead{Region} & \colhead{R.A.} &
\colhead{Decl.} & \colhead{Group} &  \colhead{$\log(T_{eff})$} &
\colhead{$M$} & \colhead{$\log(L_{bol})$} &
\colhead{$\log(L_{X})$} & \colhead{$\log(L_{X,up})$} & \colhead{F1} & \colhead{F2}\\
\colhead{} & \colhead{(deg)} &  \colhead{(deg)} & \colhead{} & \colhead{(K)} & \colhead{($M_{\odot}$)} & \colhead{($L_{\odot}$)} & \colhead{(erg s$^{-1}$)} & \colhead{(erg s$^{-1}$)} & \colhead{} & \colhead{}\\
\colhead{(1)} & \colhead{(2)} & \colhead{(3)} & \colhead{(4)} & \colhead{(5)} & \colhead{(6)} & \colhead{(7)} & \colhead{(8)} & \colhead{(9)} & \colhead{(10)} & \colhead{(11)}
}
%\decimalcolnumbers
\startdata
NGC 6242 & 253.922868 & -39.264972 & 1 & 3.69 & 0.77 & -0.55 & \nodata & 30.11 & 1 & 0\\
NGC 6242 & 253.733948 & -39.321718 & 1 & 3.72 & 0.86 & -0.36 & \nodata & 30.36 & 1 & 0\\
NGC 6242 & 253.757932 & -39.318616 & 1 & 3.75 & 0.95 & -0.23 & \nodata & 30.04 & 1 & 0\\
NGC 6242 & 253.746098 & -39.312057 & 1 & 4.01 & 2.25 & 1.44 & \nodata & 30.04 & 1 & 0\\
NGC 6242 & 253.708724 & -39.317501 & 1 & 3.83 & 1.35 & 0.50 & \nodata & 30.36 & 1 & 1\\
NGC 6242 & 253.671724 & -39.421617 & 5 & 3.93 & 1.75 & 1.00 & 30.61 & \nodata & 1 & 1\\
NGC 6242 & 253.674725 & -39.389450 & 6 & 3.55 & 0.60 & -1.15 & 30.47 & \nodata & 1 & 0\\
NGC 6242 & 253.688589 & -39.493388 & 6 & 3.60 & 0.69 & -0.90 & 29.99 & \nodata & 1 & 0\\
NGC 6242 & 253.701925 & -39.493684 & 6 & 3.96 & 1.90 & 1.15 & 30.23 & \nodata & 1 & 0\\
NGC 6242 & 253.711856 & -39.598526 & 6 & 3.57 & 0.65 & -1.04 & 29.97 & \nodata & 1 & 0\\
\enddata
\tablecomments{This table is available in its entirety (2,142 cluster members) in the machine-readable form in the online journal. Only sources from Groups 1, 2, 5, and 6 with available mass estimates are presented. Out of the total 2,142 cluster members, 1,531 are located in the central parts of the clusters (Column 10 flag $=0$), and are included in the science analysis of the X-ray-mass-age relations. Column 1: Cluster name. Columns 2-3: Gaia right ascension and declination (in decimal degrees) for epoch J2000.0. Column 4: Source class: groups 1, 2 (non-X-ray stars), and groups 5, 6 (X-ray stars). Columns 5-7: Stellar effective temperature, mass, and bolometric luminosity derived from the Gaia color-magnitude diagrams. Columns 8-9: X-ray luminosity (for X-ray members; groups 5 and 6) and upper limits to X-ray luminosity (for non-X-ray members; groups 1 and 2). Column 10: A flag indicating whether the star is located inside ($= 0$) or outside ($= 1$) the central region of the cluster ($R<7.5\arcmin$); this flag is used to determine inclusion in, or exclusion from, the activity-mass-age science analysis. Column 11: A flag indicating sources with inconsistency ($=1$) between the Gaia $G$-band flux and the combined BP$+$RP fluxes.}
\end{deluxetable*}

As a final step to minimize contamination from unrelated field stars, we apply a photometric criterion based on the location of candidate members in the Gaia color–magnitude diagram (CMD), specifically the $M_G$ vs.\ $G_{BP} - G_{RP}$ plane. We focus this filtering on the most reliable membership categories --- Groups 1, 2, 5, and 6 --- which together form a well-defined pre-main sequence locus on the CMD. A local polynomial regression curve is fitted to these groups to trace the cluster sequence\footnote{The fit is performed using the {\it locfit} function \citep{Loader99, Loader20} in R \citep{RCoreTeam20}.}. Fainter sources that deviate significantly toward bluer colors --- by more than $3-5$ times the 68\% confidence interval in $G_{BP} - G_{RP}$ --- from this empirical sequence are excluded from the final membership list. These outliers are likely field contaminants.

Additionally, we account for instrumental limitations in the {\it Chandra} data. The sensitivity of the ACIS-I array declines with increasing off-axis angle, leading to a notable drop in the detection efficiency of faint sources toward the edges of the field. To ensure a uniformly selected sample for downstream analysis of activity-age-mass relationships, we restrict our working member list to stars located within a $7.5\arcmin$ radius of the cluster center. Candidates beyond this boundary are excluded from the activity analysis described in later sections.

Figures~\ref{fig:mem_NGC6242} (a,b) show the distribution of final likely cluster members compared to non-members in a three-dimensional space defined by: (1) stellar proper motion distance from the cluster's proper motion centroid, (2) parallax, and (3) photometric offset from the cluster's empirical color-magnitude sequence.  This comparison is shown for NGC 6242, with analogous figures for the other clusters provided in the Appendix.

The candidate members from Groups 1, 2, 5, and 6 are shown in red (non-X-ray) and green (X-ray), while likely non-members --- defined as the remaining Gaia or Gaia–{\it Chandra} sources with astrometric and photometric data --- are shown in grey. Two panels, (a) and (b), present rotated views of the 3D proper motion-parallax-CMD space to aid visualization of the clustering.

To provide additional context, brown points mark members outside the {\it Chandra} fields identified by \citet{Cantat-Gaudin2020} in their Gaia-based survey of Galactic clusters. The cluster members (red, green, and brown) clearly occupy a compact and coherent region in this multidimensional space, in contrast to the more scattered distribution of unrelated Galactic field stars.

The final Group 1, 2, 5, and 6 cluster members, along with their basic stellar properties, are listed in Table~\ref{tab:stellar_props_open_clusters}. These properties --- including approximate effective temperatures, stellar masses, bolometric luminosities, X-ray luminosities for detected sources, and X-ray upper limits for non-detections --- are derived in the following sections.

\section{Cluster and Member Characterization for Five $\sim 45$--100~Myr Clusters} \label{sec:cluster_member_properties}

Cluster reddening and age estimates were derived by comparing theoretical isochrones with the observed Gaia color-magnitude diagrams. PARSEC v1.2S evolutionary tracks \citep{Bressan12, Chen14} were reddened using the extinction law from \citet{Luhman2020}, generating a grid of isochrones spanning a range of extinctions and stellar ages. These model sequences were then fitted to the Gaia photometry of cluster members, consisting of (1) Groups 1, 2, 5, and 6 identified in this study (excluding, for the purposes of this CMD fitting only, those members with highly uncertain Gaia colors, defined as $\sigma(G_{\mathrm{BP}} - G_{\mathrm{RP}}) > 0.08$ mag), and (2) additional members outside the {\it Chandra} fields from \citet{Cantat-Gaudin2020}. The fitting was performed using a chi-squared minimization procedure, weighted by the photometric uncertainties in the $G_{\mathrm{BP}} - G_{\mathrm{RP}}$ color.

Because stars of different masses respond differently to reddening and aging effects, we leveraged this property in the fitting: the positions of the higher-mass members ($M \gtrsim 2$~M$_{\odot}$) are more responsive to extinction whereas the locus of lower-mass stars ($M \lesssim 1$~M$_{\odot}$) more strongly reflects changes in cluster age. Accordingly, extinction values were primarily constrained using the upper main sequence while ages were tuned to reproduce the empirical low-mass sequence. The plausible age ranges for each cluster were selected so that the bounding PARSEC isochrones closely encompassed the empirical sequence traced by the lower-mass stars, as defined by the smooth {\it locfit} polynomial regression curve on the Gaia color-magnitude diagram.

To address concerns about the susceptibility of Gaia BP photometry to crowding, faint-source biases, and background contamination, we applied the $C^{*}$-based prescription of \citet{Riello2021} to identify sources whose BP$+$RP fluxes may be inconsistent with their $G$-band flux. Among the 2142 members of the five $<100$~Myr clusters, fewer than 6\% exhibit possible BP-related issues --- specifically, sources with corrected flux excess values satisfying $|C^{*}| > 5\,\sigma_{C^{*}}$ \citep{Getman2022}. These stars are flagged in Column 11 of Table~\ref{tab:stellar_props_open_clusters}. A similar fraction was reported for younger clusters by \citet{Getman2022}. We verified that excluding these stars from the CMD fits does not affect our cluster extinction or age estimates.

For NGC 6242, Figure~\ref{fig:mem_NGC6242}(c) displays the resulting color-magnitude diagram with the best-fit isochrone (red), bounding age isochrones (cyan and blue), and the empirical cluster sequence derived via {\it locfit} (green) overlaid on the observed Gaia photometry of cluster members (black points with grey horizontal error bars representing uncertainties in $G_{BP} - G_{RP}$). CMDs for the remaining clusters are provided in the Appendix. 

Table~\ref{tab:cluster_props} summarizes the number of identified Group 1, 2, 5, and 6 cluster members, along with the derived properties of each cluster, including distance, extinction, and age.

Stellar effective temperatures, bolometric luminosities, and masses for individual cluster members were derived by comparing their observed $G$-band absolute magnitudes to the predictions from the best-fitting reddened PARSEC 1.2S isochrone models. This approach ensures that the inferred stellar properties are largely insensitive to potential issues in Gaia color photometry that may affect a small subset of stars. The resulting stellar parameters are listed in Table~\ref{tab:stellar_props_open_clusters}.

With stellar masses in hand, we examine the ensemble mass distribution to assess the Initial Mass Function (IMF) for each cluster. This serves both as a consistency check --- ensuring that our data and methods yield a realistic IMF shape --- and as a way to evaluate the mass completeness of the membership sample and of the full stellar population of the clusters down to $0.1$~M$_{\odot}$. IMF histograms for NGC~6242 and the other clusters are presented in panel (d) of Figure~\ref{fig:mem_NGC6242} and in the corresponding Appendix figures. Each observed mass function is modeled using the analytic IMF formulation from \citet{Maschberger2013}, which smoothly connects a power-law high-mass slope with a log-normal turnover at lower masses. In most clusters, the IMF peaks below $0.75$~M$_{\odot}$, indicating mass completeness well below this threshold. In NGC~2353, however, the peak lies closer to $0.75$~M$_{\odot}$, guiding our choice to adopt $0.75$~M$_{\odot}$ as a completeness limit for the X-ray activity-age-mass analysis presented in later sections.

Finally, these mass distributions imply that our identified Group 1, 2, 5, and 6 members account for roughly 28\% of the total stellar population expected within the {\it Chandra} ACIS-I fields, based on extrapolation of the fitted IMF down to $0.1$~M$_\odot$.

\begin{deluxetable*}{ccccccccccc}
%\tablenum{2}
\tabletypesize{\tiny}
\tablecaption{X-ray Spectral Fits Of Stacked Data  \label{tab:xspec_fits}}
\tablewidth{0pt}
\tablehead{
\colhead{Cluster} & 
\colhead{$NC$} & \colhead{$\chi^{2}_\nu$} & \colhead{dof} & \colhead{$N_{H}$} & \colhead{$kT_2$} & \colhead{$EM_{1}$} & \colhead{$EM_2$} &  \colhead{$F_{tc,erg}$} & \colhead{$CF_{XSPEC}$} &  \colhead{$CF_{AE}$}\\
\colhead{} & 
\colhead{} & \colhead{} & \colhead{} & \colhead{$10^{22}$} & \colhead{} & \colhead{$10^{52}$} & \colhead{$10^{52}$} &  \colhead{$10^{-15}$}\\
\colhead{} &  \colhead{(cnts)} & \colhead{} & \colhead{} & \colhead{(cm$^{-2}$)} & \colhead{(keV)} &  \colhead{(cm$^{-3}$)} & \colhead{(cm$^{-3}$)} & \colhead{(erg~s$^{-1}$~cm$^{-2}$)} & \colhead{(keV/photon)} & \colhead{(keV/photon)}\\
\colhead{(1)} & \colhead{(2)} & \colhead{(3)} & \colhead{(4)} & \colhead{(5)} & \colhead{(6)} &
\colhead{(7)} & \colhead{(8)} &
\colhead{(9)} & \colhead{(10)} &
\colhead{(11)}
}
%\decimalcolnumbers
\startdata
NGC 6242 & 9176 & 1.2 & 176 & 0.25 & $3.0 \pm 0.2$ & $5.08 \pm 0.24$ & $3.05 \pm 0.15$ & $4.76 \pm 0.11$ & $2.40 \pm 0.08$ & $5.16 \pm 0.17$\\
Trumpler 3& 3101 & 1.0 & 71 & 0.19 & $2.5 \pm 0.3$ & $2.81 \pm 0.24$ & $1.77 \pm 0.17$ & $9.25 \pm 0.31$ & $1.87 \pm 0.09$ & $4.08 \pm 0.19$ \\
NGC 6204 & 3106 & 1.1 & 82 & 0.28 & $3.3 \pm 0.4$ & $3.08 \pm 0.26$ & $1.89 \pm 0.17$ & $3.34 \pm 0.12$ & $2.68 \pm 0.14$ & $6.48 \pm 0.34$ \\
NGC 2353 & 1975 & 1.2 & 56 & 0.08 & $6.3 \pm 2.6$ & $3.20 \pm 0.22$ & $0.87 \pm 0.12$ & $2.85 \pm 0.12$ & $1.36 \pm 0.08$ & $4.23 \pm 0.25$\\
NGC 2301 & 1800 & 1.5 & 46 & 0.05 & $1.7 \pm 0.2$ & $1.24 \pm 0.23$ & $1.88 \pm 0.19$ & $3.45 \pm 0.16$& $1.20 \pm 0.08$ & $2.79 \pm 0.18$\\
\enddata
\tablecomments{Column 1: Cluster name. Column 2: Total number of net (background-corrected) X-ray counts in a spectrum. Columns 3-4: Reduced $\chi^{2}$ for the overall spectral fit and degrees of freedom. Column 5: Fixed value of cluster's X-ray column density.  Column 6: Inferred temperature of the hot plasma component and its 1~$\sigma$ error. Columns 7-8: Inferred emission measures and their 1~$\sigma$ errors for each plasma component. Column 9: Inferred absorption corrected incident X-ray flux in the (0.5--8)~keV band. Columns 10-11: Two variants of conversion factors. The first is obtained directly from the XSPEC model fits as the unabsorbed flux in erg~cm$^{-2}$~s$^{-1}$ ($F_{tc,erg}$) divided by the absorbed flux in photons~cm$^{-2}$~s$^{-1}$ ($F_{t,phot}$). The second variant --- ultimately applied in our analyses --- is the ratio of $F_{tc,erg}$ to the sum of the $\it ACIS$-Extract–derived photometric fluxes ($\sum PFlux_i$) for all stars included in the spectral stacking, further multiplied by the ratio of the exposure time of the stacked spectrum to the median effective {\it Chandra} exposure of the participating stars ($t_{\mathrm{XSPEC}}/t_{\mathrm{Chandra,eff}}$).} 
\end{deluxetable*}

\section{X-RAY LUMINOSITIES AND UPPER LIMITS for Five $\sim 45$--100~Myr Clusters}
\label{sec:Xray_luminosities}

For all X-ray detected sources, Table~\ref{tab:xray_photometry} in the Appendix lists the apparent X-ray photometric fluxes ($PFlux$) calculated by {\it ACIS Extract}. To convert these fluxes to intrinsic X-ray luminosities ($L_X$), we derive $PFlux$-to-$L_X$ conversion factors based on stacked spectra constructed from numerous Group 5 and Group 6 X-ray-detected cluster members, and on the spectral model fits applied to those data. To avoid bias from the spectral shapes of very bright individual stars, the two brightest sources ($C_{net} > 500$ net counts) are excluded from the stacks.

The stacked spectra are processed and grouped following the procedures detailed in the {\it ACIS Extract} User Guide\footnote{\url{https://sites.psu.edu/acisextractandtarasoftware/}}. Spectral fitting is performed in XSPEC version 12.14.0h \citep{Arnaud1996}, using $\chi^2$ statistics and a two-temperature optically thin thermal plasma model subject to gas absorption, $tbabs \times (apec+apec)$ \citep{Smith2001,Wilms2000}. Cluster distances from Table~\ref{tab:cluster_props} (based on Gaia data) are used to convert fluxes to luminosities. Stellar X-ray luminosities of individual stars are then estimated from their relative contributions to the total observed photometric flux of the stacked spectrum.

In our spectral analysis, the hydrogen column density ($N_H$) is fixed according to the mean cluster extinction (Table~\ref{tab:cluster_props}), adopting the gas-to-dust ratio from \citet{Zhu2017}. The coronal metal abundance is fixed at 0.3 times the solar value, as commonly used for young and active stars \citep{Getman05,Briggs2003}.

We employ a two-temperature plasma model to represent different coronal structures, which may be heated by distinct mechanisms. The soft plasma component is fixed at $kT_1 = 0.7$ keV, consistent with plasma temperatures of $8-10$ MK observed in young or active stars. This component likely arises from numerous small-scale fundamental coronal structures and may be powered by a combination of magnetic reconnection and Alfv\'{e}n wave heating processes \citep{Preibisch05,Sanz-Forcada2003,Getman2025}. 

In contrast, the hotter plasma component ($kT_2$) is left free in the fit, along with the emission measures, $EM_1$ and $EM_2$. This hotter component is associated with giant coronal active regions and vertically extended coronal loop structures that are sites of large-scale magnetic reconnection events, which dominate the flare energy output in younger or magnetically active stars \citep{Getman2021b,Getman2025}.

The resulting {\it Chandra} spectrum for NGC 6242 is shown in Figure~\ref{fig:mem_NGC6242}(e), with analogous figures provided in the Appendix for the other clusters. Table~\ref{tab:xspec_fits} summarizes the spectral fitting results for each stacked spectrum. It includes the total number of net counts, fit quality, column density, hot plasma temperature, emission measures, and the total absorption-corrected X-ray flux of the stacked spectrum ($F_{tc,erg}$). 

The latter is computed in XSPEC using the {\it flux} command under the assumption of zero absorption. XSPEC also provides the corresponding photon fluxes in units of photons s$^{-1}$ cm$^{-2}$ (apparent and absorption-corrected). 

We cannot use the correction factor provided by XSPEC, defined as the simple ratio of the unabsorbed flux in erg~cm$^{-2}$~s$^{-1}$ to the absorbed flux in photons~cm$^{-2}$~s$^{-1}$, because the {\it ACIS-Extract}-derived source photometric fluxes ($PFlux$) are good, but not perfect, surrogates for the source's absorbed flux falling on the telescope aperture (see the detailed explanation in the {\it ACIS-Extract} manual). 

The correct approach is to use the XSPEC model fit result for the unabsorbed flux $F_{tc,erg}$ in combination with the sum of the {\it ACIS-Extract}-derived photometric fluxes ($\sum PFlux_i$) for the $N$ stars included in the spectral stacking. Additionally, because fluxes returned from XSPEC for {\it ACIS-Extract}-constructed stacked spectra are averaged over the $N$ stars, the correction factor must also be multiplied by the ratio of the total exposure of the stacked spectrum ($t_{\mathrm{XSPEC}}$) to the median effective {\it Chandra} exposure of the participating stars ($t_{\mathrm{Chandra,eff}}$):

\[
CF_{AE} = \frac{F_{tc,erg}}{\sum PFlux_i} \times \frac{t_{\mathrm{XSPEC}}}{t_{\mathrm{Chandra,eff}}}.
\]

See Table 4 for the values of both the naive correction factor ($CF_{XSPEC}$) and the more accurate factor used in this study ($CF_{AE}$). We use $CF_{AE}$ to calibrate individual {\it ACIS Extract}-based $PFlux$ measurements, allowing us to estimate the intrinsic X-ray luminosities of individual Group 5 and 6 cluster members.

This X-ray spectral analysis, conducted on the combined young stellar samples across different mass ranges, reveals no significant variation in the spectral hardness among the five clusters. For instance, while NGC 2301 exhibits a modestly lower hot plasma temperature of $kT_2 = 1.7 \pm 0.2$~keV compared to the other clusters, this is offset by a relatively higher emission measure ratio of $EM_2/EM_1 = 1.5 \pm 0.3$, in contrast to the typical $EM_2/EM_1 \sim 0.6 \pm 0.08$ seen in three of the remaining four clusters. However, as discussed in the following sections, a more detailed spectral analysis of carefully selected mass-stratified and mass-complete young stellar samples reveals noteworthy differences in the X-ray spectral properties of young stars across specific mass ranges.

For non-X-ray-detected cluster members in Groups 1 and 2, upper limits on their X-ray luminosities are estimated based on the faint-source sensitivity limits appropriate for each cluster. These limits are corrected for the decline in {\it Chandra} point source sensitivity at larger off-axis angles \citep{Feigelson2002}. In Figure~\ref{fig:mem_NGC6242} (panel f) for NGC 6242 --- and in comparable figures for the other clusters presented in the Appendix --- the distribution of X-ray luminosities as a function of off-axis angle is plotted for X-ray-detected cluster members (Groups 5 and 6) and, for reference, for unrelated X-ray sources (Groups 7-10, under the unreliable assumption that they are located at the same distance as the cluster). The red curves represent locally weighted likelihood-based quadratic fits to the lower envelope of the X-ray luminosity distribution, computed using the {\it locfit.robust} function from the R CRAN {\rm locfit} package \citep{Loader99,Loader20}. These fitted curves are then used to assign upper limits to the X-ray luminosities of non-detected sources based on their angular distances from the cluster centers. Final values of X-ray luminosities and upper limits for all cluster members are compiled in Table~\ref{tab:stellar_props_open_clusters}.

\section{Extension to Older Clusters} \label{sec:3more_older_clusters}

In the following section (\S~\ref{sec:resuts}), we show that the {\it Chandra}-Gaia-based median X-ray luminosities of sub-solar-mass stars ($0.75$--$0.9$M${_{\odot}}$) in $7$--$100$~Myr-old open clusters are consistent with the predictions of the semi-empirical study by \citet{Johnstone2021}, whereas solar-mass stars ($0.9$-$1.2$~M$_{\odot}$) in the $\sim 100$~Myr-old clusters NGC~2353 and NGC~2301 exhibit X-ray luminosities well below these predictions. To further examine this discrepancy, we extend our analysis to three older clusters, NGC~6475, M37, and Praesepe, with ages $t \sim 220$--750~Myr (Table~\ref{tab:extra3cluster_props}).

\begin{deluxetable*}{lrrcrcccc}
%\tablenum{2}
\tabletypesize{\footnotesize}
\tablecaption{Three Open Clusters with Ages $> 150$~Myr\label{tab:extra3cluster_props}}
\tablewidth{0pt}
\tablehead{
\colhead{Region} & \colhead{$l$} &
\colhead{$b$} & \colhead{$N_{mem;0.75-1.2M_{\odot}}$} &  \colhead{$D$} & \colhead{$A_{V}$} & \colhead{Age} & \colhead{$N_{>0.1M_{\odot}}$} & \colhead{$M_{lim}$}\\
\colhead{} & \colhead{(deg)} &  \colhead{(deg)} & & \colhead{(pc)} & \colhead{(mag)} & \colhead{(Myr)} & \colhead{} & \colhead{(M$_{\odot}$)}\\
\colhead{(1)} & \colhead{(2)} & \colhead{(3)} & \colhead{(4)} & \colhead{(5)} & \colhead{(6)} & \colhead{(7)} & \colhead{(8)} & \colhead{(9)}
}
%\decimalcolnumbers
\startdata
NGC 6475 & 355.8359 & -4.5280 & 268 (55) & $276 \pm 1$ & 0.30 & 220 (150 - 300) & 2972 & $<0.75$\\
M37 & 177.6484 & 3.0890 & 496 (124) & $1483 \pm 3$ &     0.80 & 500 (400 - 700) & 6686 & $>1.2$\\
Praesepe & 205.9158 & 32.4700 & 119 (46) & $183 \pm 1$ & 0.16 & 750 (600 - 900) & 1194 & $<0.75$\\
\enddata
\tablecomments{Column 1: Cluster name. Columns 2-3: Galactic coordinates for the cluster center in degrees. Column 4: Total number of $0.75-1.2$~M$_{\odot}$ cluster members identified  by \citet{HuntReffert2023} across the entire cluster. Values in parentheses indicate a subsample of members located in the central regions of the clusters, which is employed in the analysis of the X-ray-mass-age relations; this subsample totals 225 X-ray and non-X-ray stars (Table~\ref{tab:tab_ngc6475_m37_praesepe}). Column 5: Median cluster distance from the Sun and its 68\% bootstrap error. Column 6: Cluster-average extinction in the visual band. Column 7: Cluster age, where the best-fit value corresponds to the PARSEC PMS isochrone that most closely follows the main-sequence turn-off (MSTO) region in the Gaia CMD; values in parentheses indicate approximate lower and upper age limits, corresponding to the isochrones bracketing the MSTO. Column 8: IMF-based estimate of the intrinsic total number of cluster members down to 0.1~M$_{\odot}$. Column 9: Mass completeness limits for the cluster member samples.} 
\end{deluxetable*}

\begin{figure*}[ht!]
\plotone{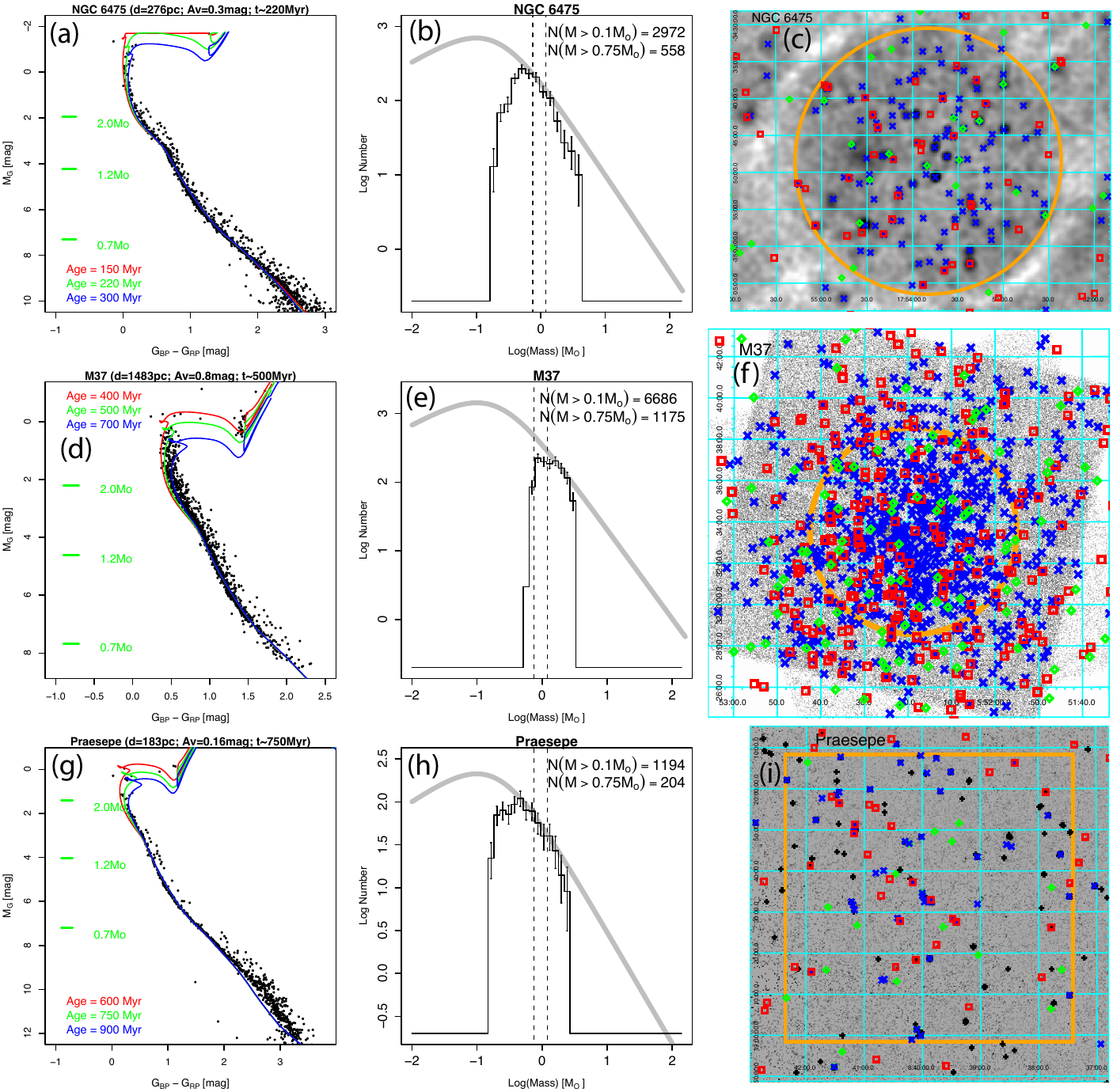}
\caption{(a, d, g) Gaia CMDs of all cluster members with $Prob > 0.5$ in the $>150$~Myr clusters NGC~6475, M37, and Praesepe \citep{HuntReffert2023}. The best-fit PARSEC isochrone to the MSTO is shown in green, with bounding isochrones surrounding the MSTO in red and blue.  (d, e, h) IMF histograms of the same cluster members (black), with the theoretical IMF from \citet{Maschberger2013} shown in grey. Dashed vertical lines mark $0.75$ and $1.2$~M$_{\odot}$, and the legend lists both the observed number of stars with $M>0.75$~M$_\odot$ and the extrapolated population down to $0.1$~M$_{\odot}$. (c, f, i) Spatial distributions of Gaia cluster members with $Prob > 0.5$ in the mass ranges $0.75-0.9$~M$_{\odot}$ (green diamonds) and $0.9-1.2$~M$_{\odot}$ (red boxes), overlaid with X-ray detections (blue $\times$) from ROSAT (NGC~6475 and Praesepe) and Chandra (M37). For Praesepe, the ROSAT PSPC catalog of \citet{Randich1995} (blue $\times$), limited to stars considered members at the time, is supplemented with additional ROSAT sources from the Second ROSAT PSPC catalog (black $+$) following Gaia-based membership updates. Only X-ray and non-X-ray members located within the orange contours are used in the X-ray temporal evolution analysis (Figure~\ref{fig:lx_vs_t_main}) and are recorded in Table~\ref{tab:tab_ngc6475_m37_praesepe}. The demarcating orange circles for NGC~6475 and M37 have radii of $R=18\arcmin$ and $5\arcmin$, respectively, while the orange square for Praesepe measures $70 \times 70$~arcmin$^2$. \label{fig:NGC6475_M37_Praesepe}}
\end{figure*}

\begin{deluxetable*}{cccccccccc}
%\tablenum{2}
\tabletypesize{\small}
\tablecaption{Sub-solar- and Solar-mass Members With and Without X-rays in Three $t>150$~Myr Clusters \label{tab:tab_ngc6475_m37_praesepe}}
\tablewidth{0pt}
\tablehead{
\colhead{Cluster} & \colhead{Id HR2023} & \colhead{R.A.} &
\colhead{Decl.} & \colhead{Prob} & \colhead{Mass} & \colhead{Gaia-X-ray Sep.} & \colhead{Id X-ray} &
\colhead{$\log(L_X)$} & \colhead{$\log(L_{X,up})$} \\
\colhead{} & \colhead{} &  \colhead{(deg)} &
\colhead{(deg)} & \colhead{} & \colhead{(M$_{\odot}$)} & \colhead{($\arcsec$ )} & \colhead{} &
\colhead{(erg~s$^{-1}$)} & \colhead{(erg~s$^{-1}$)} \\
\colhead{(1)} & \colhead{(2)} & \colhead{(3)} & \colhead{(4)} & \colhead{(5)} &
\colhead{(6)} & \colhead{(7)} & \colhead{(8)} & \colhead{(9)} & \colhead{(10)} 
}
\startdata
NGC6475 & 1048 & 268.629734 & -34.665050 & 0.966 & 1.00 & 6.0 & 116 & 29.41 & \nodata\\
NGC6475 & 1049 & 268.608917 & -34.703623 & 0.828 & 1.01 & 9.4 & 109 & 29.27 & \nodata\\
NGC6475 & 1060 & 268.558018 & -34.556302 & 0.741 & 0.84 & 133.1 & \nodata & \nodata & 29.03\\
NGC6475 & 1063 & 268.492311 & -34.629297 & 0.945 & 1.07 & 45.8 & \nodata & \nodata & 28.93\\
NGC6475 & 1065 & 268.416777 & -34.667791 & 0.677 & 0.99 & 7.7 & 66 & 29.10 & \nodata\\
NGC6475 & 1068 & 268.388041 & -34.711606 & 0.966 & 0.79 & 10.2 & 57 & 28.91 & \nodata\\
NGC6475 & 1069 & 268.451290 & -34.702931 & 0.962 & 0.92 & 119.6 & \nodata & \nodata & 28.84\\
NGC6475 & 1076 & 268.324754 & -34.625622 & 0.917 & 1.15 & 10.4 & 36 & 29.50 & \nodata\\
NGC6475 & 1077 & 268.594789 & -34.732859 & 0.852 & 1.06 & 11.9 & 105 & 29.08 & \nodata\\
NGC6475 & 1079 & 268.605729 & -34.793587 & 1.000 & 0.90 & 68.7 & \nodata & \nodata & 28.86\\
\enddata
\tablecomments{This table is published in its entirety in machine-readable form in the online journal. It lists 225 members with masses $0.75$--$1.2$~M$_{\odot}$ in NGC~6475, M37, and Praesepe, located within the orange circles and boxes shown in Figure~\ref{fig:NGC6475_M37_Praesepe}. Cluster membership is based on the Gaia open cluster study of \citet{HuntReffert2023}, while X-ray luminosities (or upper limits) are taken from ROSAT and Chandra studies \citep{Prosser1995, Randich1995, Nunez2015}. Column 1: Cluster name.  Column 2: Source sequential number from \citet{HuntReffert2023} (note that numbering restarts from 1 in each cluster). Columns 3-4: Gaia right ascension and declination (J2000.0, decimal degrees).  Column 5: Cluster membership probability from \citet{HuntReffert2023}, with only stars having $P > 0.5$ included. Column 6: Stellar mass, estimated from the Gaia CMD using PARSEC stellar evolutionary models. Column 7: Angular separation between the Gaia star and its nearest X-ray source. For ROSAT data (NGC~6475 and Praesepe), a match requires separation $<20\arcsec$; for Chandra data (M37), separation $<1\arcsec$. Column 8: X-ray source identifier: sequential number from \citet{Prosser1995, Randich1995}, IAU designation from \citet{Nunez2015}, or (for Praesepe) 2RXP designation from the Second ROSAT PSPC catalog \citep{Rosat2000}, used to complement ROSAT sources not reported by \citet{Randich1995}. Columns 9-10: X-ray luminosity or upper limit to X-ray luminosity.}
\end{deluxetable*}

ROSAT PSPC (Position Sensitive Proportional Counter; \citealt{Truemper1982}) observations of NGC~6475 and Praesepe are reported by \citet{Prosser1995} and \citet{Randich1995}, respectively, while {\it Chandra}/ACIS observations of M37 are described in \citet{Nunez2015,Nunez2016}.

Our concern is not with the reliability of the X-ray luminosity measurements in these studies, but with the pre-Gaia membership determinations, which were based solely on optical color-magnitude diagrams. These memberships have been substantially revised using Gaia astrometric and photometric data. To save time, we do not perform our own Gaia membership analysis for these three clusters; instead, we adopt the stellar member lists from the Gaia-DR3 catalog of \citet{HuntReffert2023}, who applied density-based clustering combined with machine-learning validation to produce large, homogeneous catalogs of clusters and their candidate members. Our analysis is restricted to candidate members with a cluster membership probability greater than $0.5$.

Figure~\ref{fig:NGC6475_M37_Praesepe} presents Gaia color-magnitude diagrams with best-fit PARSEC isochrones, the corresponding IMF histograms, and spatial distributions of X-ray detections (blue; additionally black for Praesepe) and Gaia cluster members (solar-mass stars in red, sub-solar-mass stars in green) overlaid on the combined low-resolution ROSAT PSPC images (NGC~6475 and Praesepe) and the {\it Chandra}/ACIS-I image (M37). Table~\ref{tab:tab_ngc6475_m37_praesepe} lists the resulting sub-solar- and solar-mass Gaia cluster members, including their X-ray luminosities or upper limits. All these cluster members are located within the orange circles and boxes shown in Figure~\ref{fig:NGC6475_M37_Praesepe}, which mark regions of high PSPC and ACIS sensitivity and high X-ray detection density. Further details are provided below.

The core of NGC~6475 (distance from the Sun  $d=276$~pc) was observed with two nearly identical ROSAT PSPC pointings, yielding a total exposure time of 47~ksec \citep{Prosser1995}. That study reported over 140 X-ray detections with luminosities in the $0.07$--$2.4$~keV band. As with other PSPC observations, the PSF broadens with increasing off-axis angle, and the detector's effective sensitivity decreases away from the optical axis. For our analysis of NGC~6475, we restrict attention to stars within the central $R<18\arcmin$ region of the detector (orange circle in Figure~\ref{fig:NGC6475_M37_Praesepe}c), which contains the vast majority of detected X-ray sources.

The nearer Praesepe cluster ($d=183$~pc) was covered by a $16$~deg$^2$ ROSAT PSPC mosaic \citep{Randich1995}. That study published a list of 68 X-ray detections and their $0.1$--$2.4$~keV luminosities (blue $\times$ in Figure~\ref{fig:NGC6475_M37_Praesepe}i), limited to sources matched with their pre-Gaia list of possible cluster members. Using the Second ROSAT PSPC Catalog \citep{Rosat2000}, we recover numerous additional X-ray detections associated with this mosaic (black $+$). Our analysis focuses on the central $70\arcmin \times 70\arcmin$ region of Praesepe, which has PSPC exposure map values of 10--28~ksec and contains the highest density of Gaia cluster members (Figure~\ref{fig:NGC6475_M37_Praesepe}i).

For both NGC~6475 and Praesepe, ROSAT detections were cross-matched with the Gaia cluster member catalog of \citet{HuntReffert2023} using a $20\arcsec$ search radius. This choice of matching radius, considerably larger than the 1\arcsec\ radius used for \textit{Chandra} (whose ACIS-I PSF FWHM is $\sim$0.5\arcsec\ on-axis), reflects the much larger ROSAT positional uncertainties: the PSPC PSF has FWHM $\sim$25\arcsec\ on-axis and broadens substantially off-axis, resulting in typical positional errors of 10--20\arcsec. A smaller radius would risk missing genuine counterparts. The 20\arcsec\ radius is consistent with previous ROSAT studies \citep[e.g.,][]{Prosser1995,Morley2001} and balances completeness with the low probability of spurious matches, particularly since we restrict the analysis to confirmed Gaia cluster members and to central PSPC regions with higher sensitivity. 

Reported X-ray luminosities were corrected for small differences between the cluster distances adopted in \citet{Prosser1995,Randich1995} and the current Gaia-based values. Using the {\it fakeit} command in XSPEC and assuming an $apec$ plasma temperature of $kT=1$~keV, we find $L_{X,0.5-8}/L_{X,0.1-2.4} = 0.74$ and apply this factor to convert ROSAT luminosities to the $0.5$--$8$~keV band. Assuming a softer plasma ($kT=0.5$~keV) yields a slightly different conversion factor of $L_{X,0.5-8}/L_{X,0.1-2.4} = 0.68$. Assuming a cooler coronal plasma with $kT = 0.5$ keV would systematically lower the inferred $L_X$ values for these older clusters by a scientifically negligible amount, approximately 0.05 dex in $\log(L_X)$, in the final X-ray evolution figures presented in the next section, \S~\ref{sec:results_temporal_evolution}. For Gaia cluster members undetected by ROSAT, X-ray luminosity upper limits were estimated from the empirical relation between $L_X$ and exposure map value (see Figure~4 in \citealt{Randich1995}).

For M37, the deep $\sim 440$~ksec {\it Chandra}-ACIS observations yielded over 770 X-ray point sources, primarily located on the ACIS-I chips \citep{Nunez2015}. The X-ray catalog was cross-matched with Gaia cluster member data using a 1$\arcsec$ search radius. For the X-ray-age analysis, we retain stars within the $R < 5\arcmin$ central region of the detector (orange circle in Figure\ref{fig:NGC6475_M37_Praesepe}f), where the fraction of X-ray non-detections is lower. Since \citet{Nunez2015} reported X-ray luminosities for only 36\% of their X-ray point sources, we re-ran {\it ACIS Extract} and applied our X-ray analysis procedures (\S \ref{sec:Xray_luminosities}) to derive X-ray luminosities and upper limits for all X-ray detected and undetected Gaia cluster members. For sources with luminosities previously reported by \citet{Nunez2015}, our results are in excellent agreement with theirs.

At these older cluster ages, the main-sequence turn-off (MSTO) region on the Gaia CMD provides the primary age diagnostic. Figures~\ref{fig:NGC6475_M37_Praesepe}a,d,g show the best-fit PARSEC isochrone to the MSTO in green, with bounding isochrones surrounding the MSTO in red and blue. The inferred Gaia ages are in excellent agreement with previously published values. A known discrepancy between observed Gaia color-magnitude sequences for very low-mass stars in older clusters and PARSEC model predictions \citep{Wang2025} is apparent in the CMD of Praesepe, but it does not affect our MSTO-based age determination. The related IMFs (Figures~\ref{fig:NGC6475_M37_Praesepe}b,e,h) indicate that the considered cluster member samples are well complete below $M_{lim} < 0.75$~M$_{\odot}$ for the nearby NGC 6475 and Praesepe clusters, but reach shallower completeness limits, M$_{lim} > 1.2$~M$_{\odot}$, for the more distant M37 cluster.

%% Please use the acknowledgment and contribution environments. This will 
%% be anonomyized when the "anonymous" style option is used. 

\begin{figure*}[ht!]
\plotone{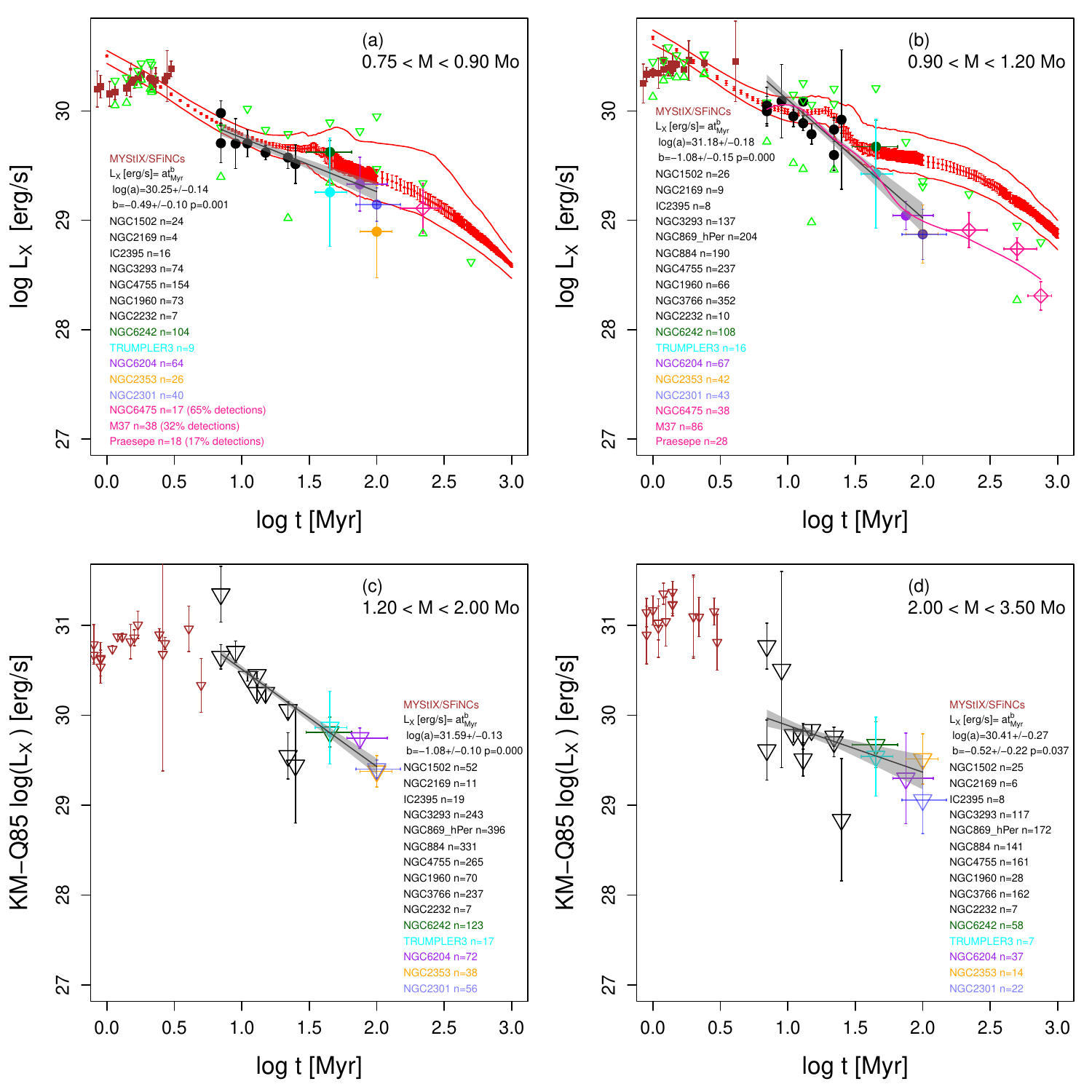}
\caption{Temporal evolution of X-ray luminosity (in the $0.5-8$~keV band) across four stellar mass strata. The corresponding mass bins are indicated in the panel legends. Younger clusters from \citet{Getman2022} are shown as brown points ($t<5$~Myr) and black points ($7-22$~Myr). The five new ($30-150$~Myr) {\it Chandra} clusters (Table~\ref{tab:cluster_props}) are marked by dark green, cyan, purple, orange, and blue points, while three older clusters ($>150$~Myr) from an extended sample with previously published X-ray detections (Table~\ref{tab:extra3cluster_props}) are represented by large pink points. The latter appear only in the first two panels. The legends list the cluster names and the total number of stars (X-ray detections and non-detections) used in the Kaplan-Meier (KM) estimation of $L_X$, as well as the best-fit parameters from weighted least squares regression fits to the relation $L_X = a \times t^b$, performed over the $7-100$~Myr baseline using the R {\it lm} function \citep{Sheather09}. The regression fits are shown as black lines with 68\% confidence intervals shaded in gray; and the small $p$-values (except for panel (d)) reported in the legends indicate strong evidence against the null hypothesis of a zero slope. In addition, for panel (b) only, the {\it locfit} best-fit (pink curve) shows a declining trend of X-ray luminosity over the extended $7-750$~Myr baseline.  (a,b) Panels show the evolution of $L_X$ for lower-mass stars. Median $L_X$ values (50\% quartiles) are shown with 68\% bootstrap confidence intervals: circles or diamonds for open clusters and boxes for younger MYStIX/SFiNCs clusters. The 25\% and 75\% quartiles, plotted only for samples with more than 15 stars and where permitted by the KM estimator, are marked by green triangles. For comparison, the model $L_X$ predictions in the $0.5-7$~keV band from \citet{Johnstone2021} are shown: red points with error bars for medians and their 68\% confidence intervals, and red lines for the interquartile range (25\%-75\%). (c,d) Panels show analogous information for higher-mass stars, but only the 85\% $L_X$ quartiles (as permitted by the KM estimator) are plotted, with their associated 68\% bootstrap confidence intervals. \label{fig:lx_vs_t_main}}
\end{figure*}

\begin{deluxetable*}{ccccccc}
%\tablenum{2}
\tabletypesize{\small}
\tablecaption{Evolution of X-ray luminosity in $7-100$~Myr-old Stars \label{tab:temporal_evolution_table}}
\tablewidth{0pt}
\tablehead{
\colhead{Mass Stratum} & \colhead{Quantity} &
\colhead{$N_{clusters}$} & \colhead{$N_{stars}$} & \colhead{$\log(a)$} & \colhead{$b$} & \colhead{$p$-val}\\
\colhead{$M_{\odot}$} & \colhead{erg~s$^{-1}$} &
\colhead{} & \colhead{} & \colhead{} & \colhead{} & \colhead{}\\
\colhead{(1)} & \colhead{(2)} & \colhead{(3)} & \colhead{(4)} & \colhead{(5)} & \colhead{(6)} &
\colhead{(7)}
}
%\decimalcolnumbers
\startdata
$ 0.75 - 0.80 $ & $L_X$ & 11 & 327 & $ 30.16 \pm 0.12 $ & $ -0.41 \pm 0.09 $ & 0.001 \\
$ 0.75 - 0.85 $ & $L_X$ & 12 & 450 & $ 30.27 \pm 0.13 $ & $ -0.50 \pm 0.10 $ & 0.000 \\
$ 0.75 - 0.90 $ & $L_X$ & 12 & 595 & $ 30.25 \pm 0.14 $ & $ -0.49 \pm 0.10 $ & 0.001 \\
$ 0.90 - 1.20 $ & $L_X$ & 15 & 1515 & $ 31.18 \pm 0.18 $ & $ -1.08 \pm 0.15 $ & 0.000 \\
$ 0.95 - 1.20 $ & $L_X$ & 15 & 1412 & $ 31.21 \pm 0.18 $ & $ -1.10 \pm 0.15 $ & 0.000 \\
$ 0.95 - 1.05 $ & $L_X$ & 15 & 730 & $ 30.85 \pm 0.23 $ & $ -0.90 \pm 0.20 $ & 0.000 \\
$ 0.95 - 1.10 $ & $L_X$ & 15 & 1013 & $ 31.14 \pm 0.21 $ & $ -1.09 \pm 0.18 $ & 0.000 \\
$ 0.95 - 1.15 $ & $L_X$ & 15 & 1236 & $ 31.13 \pm 0.18 $ & $ -1.07 \pm 0.15 $ & 0.000 \\
$ 1.20 - 2.00 $ & Q85\%~$L_X$ & 15 & 1937 & $31.59 \pm 0.13$ & $-1.08 \pm 0.10$ & 0.000 \\  
$ 2.00 - 3.50 $ & Q85\%~$L_X$ & 15 & 965 & $30.41 \pm 0.27$ & $-0.52 \pm 0.22$ & 0.037 \\  
\enddata
\tablecomments{Column 1: Mass stratum. Column 2: X-ray luminosity in the $(0.5-8)$~keV band: median for lower mass stars and 85\% quantiles for higher mass stars. Columns 3-4: Numbers of open clusters and cluster members. Columns 5-7: Results from the linear regression fits for the relations $Q=a \times t_{Myr}^b$ within the $7-100$~Myr age range, where Q is one of the above stellar quantities. The results include the intercept and slope with 68\% errors, and $p$-value for the hypothesis of zero slope. See associated Figures~\ref{fig:lx_vs_t_main} and~\ref{fig:lx_vs_t_appendix}.
}
\end{deluxetable*}

\section{Results} \label{sec:resuts}

\subsection{X-ray Emission Evolution in Young Stars: Dependence on Mass and Age} \label{sec:results_temporal_evolution}

According to PARSEC evolutionary stellar models \citep[see Figure~1 in][]{Getman2022}, some lower-mass (0.75--0.9~M$_{\odot}$) stellar members of the newly added $t>30$~Myr clusters are still evolving along Henyey tracks in the Hertzsprung-Russell diagram, whereas most stars with $>0.9$~M$_{\odot}$ have already reached the zero-age main sequence (ZAMS). In the mass range of $0.75$-$1.3$~M$_\odot$, the radiative core grows during the pre-main sequence Henyey phase and continues to expand, though more gradually, during and after the ZAMS stage \citep{Iben65}. At around $1.3$-$1.5$~M$_{\odot}$, the convective envelope vanishes as the star reaches the ZAMS and the envelope becomes fully radiative. At even higher masses, stars develop a small convective core --- driven by high energy generation via the CNO cycle --- surrounded by a large radiative envelope \citep{Hansen2004}. This ongoing internal structural evolution can significantly impact internal magnetic dynamo efficiency, surface magnetic activity, rotational behavior, and angular momentum loss.

By adding five $\sim 45$--100~Myr and three $\sim 220$--750~Myr open clusters analyzed in this study to the sample of young stellar clusters from \citet{Getman2022}, we extend the temporal baseline of the mass-stratified X-ray emission dataset from $<25$~Myr to $\sim 750$~Myr. In our current analysis, the dataset is limited to stars with masses $>0.75$~M$_{\odot}$, corresponding to the IMF-based mass completeness limit for Gaia- and X-ray-identified members of seven of the eight new clusters (Tables~\ref{tab:cluster_props} and \ref{tab:extra3cluster_props}). The dataset for the old and more distant cluster M37 remains incomplete below $1.2$~M$_{\odot}$.

To construct mass- and age-stratified X-ray luminosity functions, we follow \citet{Getman2022} in using the nonparametric Kaplan–Meier estimator \citep[KM;][]{KaplanMeier58}, which includes both X-ray detections and non-detections. Confidence intervals are based on bootstrap resampling. The KM statistics are computed using the {\it survival} package in R CRAN \citep{Therneau20}.

Figure~\ref{fig:lx_vs_t_main} shows the resulting mass-stratified evolution of X-ray emission in young stars over an expanded age range: $<1$ to $\sim 750$~Myr for lower-mass stars ($0.75$--$1.2$~M$_{\odot}$) and $<1$ to $\sim 100$~Myr for higher-mass stars ($>1.2$~M$_{\odot}$). For the lower-mass range ($<1.2$~M$_{\odot}$), fewer than 50\% of cluster members are X-ray non-detections, allowing estimation of both the median (50\%) and upper quartile (75\%) $L_X$ levels; however, the fraction of non-detections is already too high to permit estimation of lower (25\%) quartiles. For the higher-mass range ($>1.2$~M$_{\odot}$), the proportion of non-detections is so high that only the 85\% upper quartiles can be reliably derived.

Our updated analysis in Figure~\ref{fig:lx_vs_t_main}, incorporating eight older clusters, yields several key findings:  
\begin{enumerate}
    \item The decay of X-ray luminosity from 7 to 100 Myr is dramatically slower in lower-mass stars (0.75--0.9~M$_{\odot}$) compared to solar-mass stars (0.9--1.2~M$_{\odot}$). Using weighted least-squares fits to power law relation $L_X \propto t^b$, we find $b = -0.5 \pm 0.01$ for the lower-mass stratum and $b = -1.1 \pm 0.15$ for the solar-mass stratum (panels a and b). Table~\ref{tab:temporal_evolution_table} in this section and Figure~\ref{fig:lx_vs_t_appendix} in the Appendix demonstrate that there are no mass-stratum selection biases, and that the inferred distributions of $L_X$ for other mass strata within the $0.75$--$1.2$~M$_{\odot}$ range exhibit similar temporal evolution patterns.
    
An alternative interpretation of panel (b) is also possible: the X-ray luminosity decay of solar-mass stars is roughly as slow as that of sub-solar-mass stars down to $t \sim 45$~Myr, after which the decay of solar-mass stars accelerates markedly between 45 and 100~Myr.
    
    \item Incorporating data from the three older clusters ($t \sim 220$--750~Myr) yields the following results (panels a and b). For sub-solar mass stars ($0.75$--$0.9$~M$_{\odot}$), the $L_X \propto t^{-0.5}$ relation derived for the 7--100 Myr range remains valid at least to $t \sim 220$~Myr. At older ages (M37 and Praesepe), the X-ray census is dominated by non-detections, permitting only an estimate of the 75th percentile $L_X$ for M37. This suggests a possible deviation from a single power-law decay, though additional data are required to confirm such divergence. For solar-mass stars ($0.9$--$1.2$~M$_{\odot}$), the $L_X$-$t$ relation beyond $t > 150$~Myr appears visually shallower than the $L_X \propto t^{-1.1}$ established at 7--100~Myr, as indicated by the {\it locfit} trend (pink curve). Robust constraints on the shape of the long-term decay of $L_X$ for solar analogs will require more sensitive observations of more clusters in the 100--1000 Myr age range.
    \item Our purely empirical X-ray and Gaia-based results are consistent with the semi-empirical predictions of \citet{Johnstone2021} for $0.75$--$0.9$~M$_{\odot}$ stars, but reveal systematically lower X-ray emission levels for solar-mass stars at later ages. In particular, across the $100$--$750$~Myr range, the empirical median $L_X$ values fall below those predicted by \citet{Johnstone2021}, while the empirical 75th-percentile $L_X$ values lie close to the 25th-percentile levels of their model (panels a and b).

    Furthermore, independent observations of $\sim 100$~Myr-old clusters NGC~2516 and the Pleiades with the softer energy-response telescopes XMM and ROSAT \citep{Pillitteri2006,Nunez2016} show that the median X-ray luminosities of G-type stars (corresponding to our $0.9$--$1.2$~M$_{\odot}$ sample at these ages) lie near $\log(L_X) \sim 29$~erg~s$^{-1}$ --- consistent with our {\it Chandra} findings but also systematically lower by $\sim 0.5$~dex than the predictions of \citet{Johnstone2021} (panel b). This supports the interpretation that the steep decline in $L_X$ with age among solar-mass stars at $\sim 100$~Myr is a genuine astrophysical trend rather than an instrumental bias arising from {\it Chandra}'s reduced soft X-ray response.
    \item For higher-mass stars, that evolve more rapidly onto the ZAMS, the 85\% quartiles show a high level before $\sim 10$~Myr followed by a steep decline in X-ray luminosity for $1.2$--$2.0$~M$_{\odot}$ stars, and a persistently low, nearly flat emission level beyond 10~Myr for even more massive ($2.0$--$3.5$~M$_{\odot}$) stars (panels c and d).
\end{enumerate}

We also assess the potential impact of binarity on the X-ray luminosities. In particular, equal-mass binaries form a distinct ``binary sequence'' above the single-star main sequence in the Gaia CMDs for the studied open clusters (Figures~\ref{fig:mem_NGC6242}, \ref{fig:NGC6475_M37_Praesepe}, \ref{fig:mem_TRUMPLER3}–\ref{fig:mem_NGC2301}) and can be readily identified. For the combined sample of solar-mass members in the 100~Myr-old clusters NGC~2301 and NGC~2353, we performed a sanity-check analysis by removing these equal-mass binary candidates and recalculating the Kaplan–Meier estimators of the X-ray luminosity functions. The results show no significant differences in $L_X$ distributions with and without the binary candidates\footnote{For the original sample retaining equal-mass binaries ($N=85$ stars), the KM estimators yield a median $\log(L_X) = 28.87 \pm 0.19$~erg~s$^{-1}$ and 75\% quantile $\log(L_X) = 29.33$~erg~s$^{-1}$. After removing equal-mass binaries ($N=71$ stars), the corresponding KM estimators are: median $\log(L_X) = 28.77 \pm 0.18$~erg~s$^{-1}$ and 75\% quantile $\log(L_X) = 29.28$~erg~s$^{-1}$.}.

%%% ABOUT UNRESOLVED BINARIES We do not attempt to correct for the presence of unresolved binaries, which can bias both CMD-based mass stratification and X-ray luminosity estimates. Equal-mass binaries may appear overluminous in both the CMD and $L_X$ distributions, slightly inflating scatter and shifting some lower-mass stars into higher-mass bins. While these effects may modestly bias absolute normalizations, the overall age-dependent decay trends are expected to remain robust.

\begin{figure*}[ht!]
\plotone{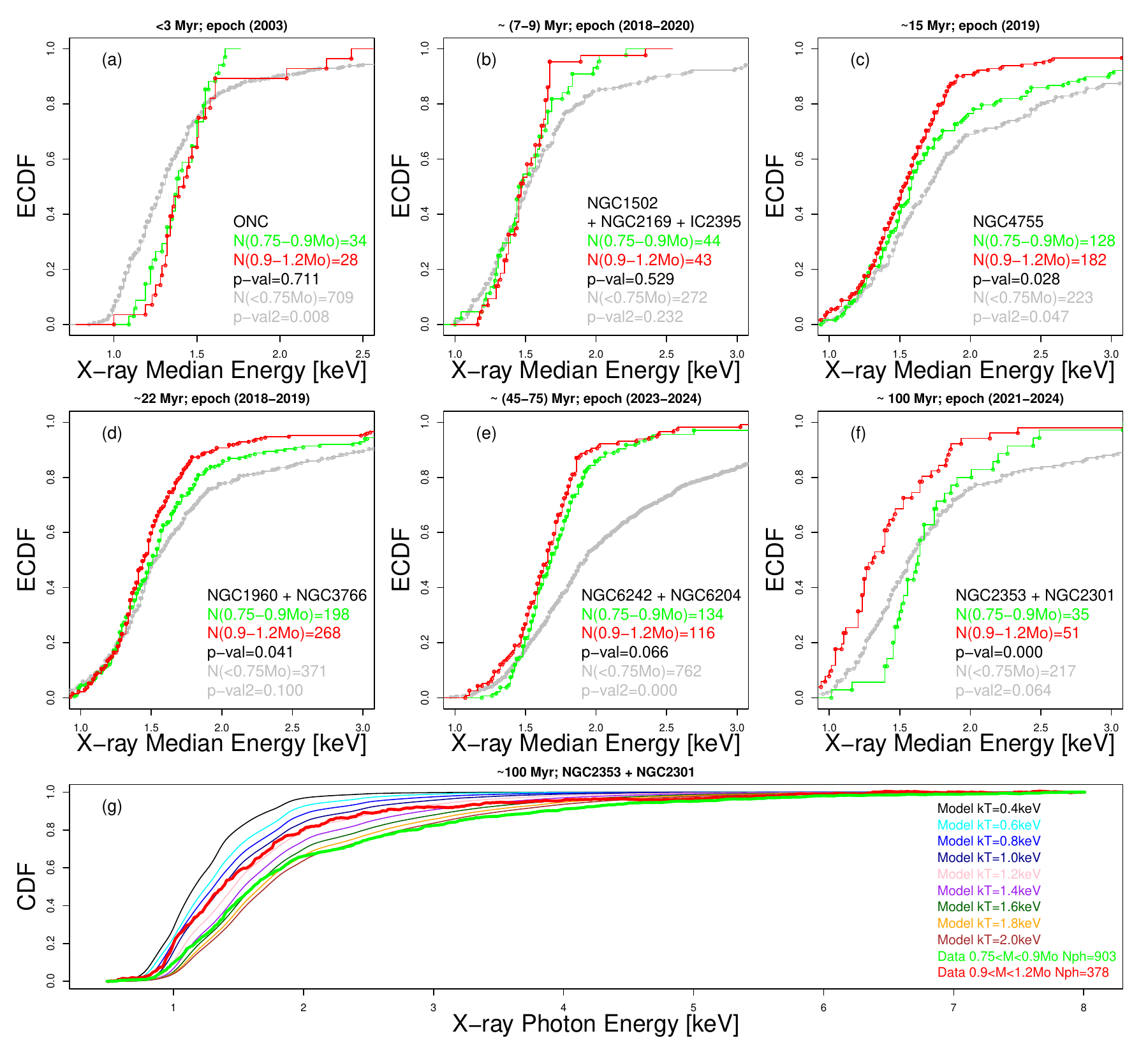}
\caption{(a-f) Empirical cumulative distribution functions (ECDFs) of source' X-ray median photon energies, stratified by stellar mass and stacked for clusters of similar ages and observation epochs. Stellar mass groups are color-coded: solar-mass (red), sub-solar (green), and lower-mass (gray). Figure legends indicate the number of stars in each group, along with $p$-values from Anderson-Darling two-sample tests (using the {\it ad.test} function from the R package {\it kSamples}) for pairwise comparisons between sub-solar and solar-mass samples (black text), and sub-solar and lower-mass samples (gray text). (g) For the oldest clusters, NGC~2353 and NGC~2301 ($\sim 100$~Myr), ECDFs of the observed, background-subtracted photon energies are shown for solar-mass (red) and sub-solar (green) stars. These distributions are overlaid with simulated CDFs from one-temperature coronal plasma models with varying plasma temperatures. The figure legend specifies the plasma temperatures corresponding to each model curve, as well as the total number of net counts used to construct the observed photon energy distributions. \label{fig:me_evolution}}
\end{figure*}

\begin{figure*}[ht!]
\plotone{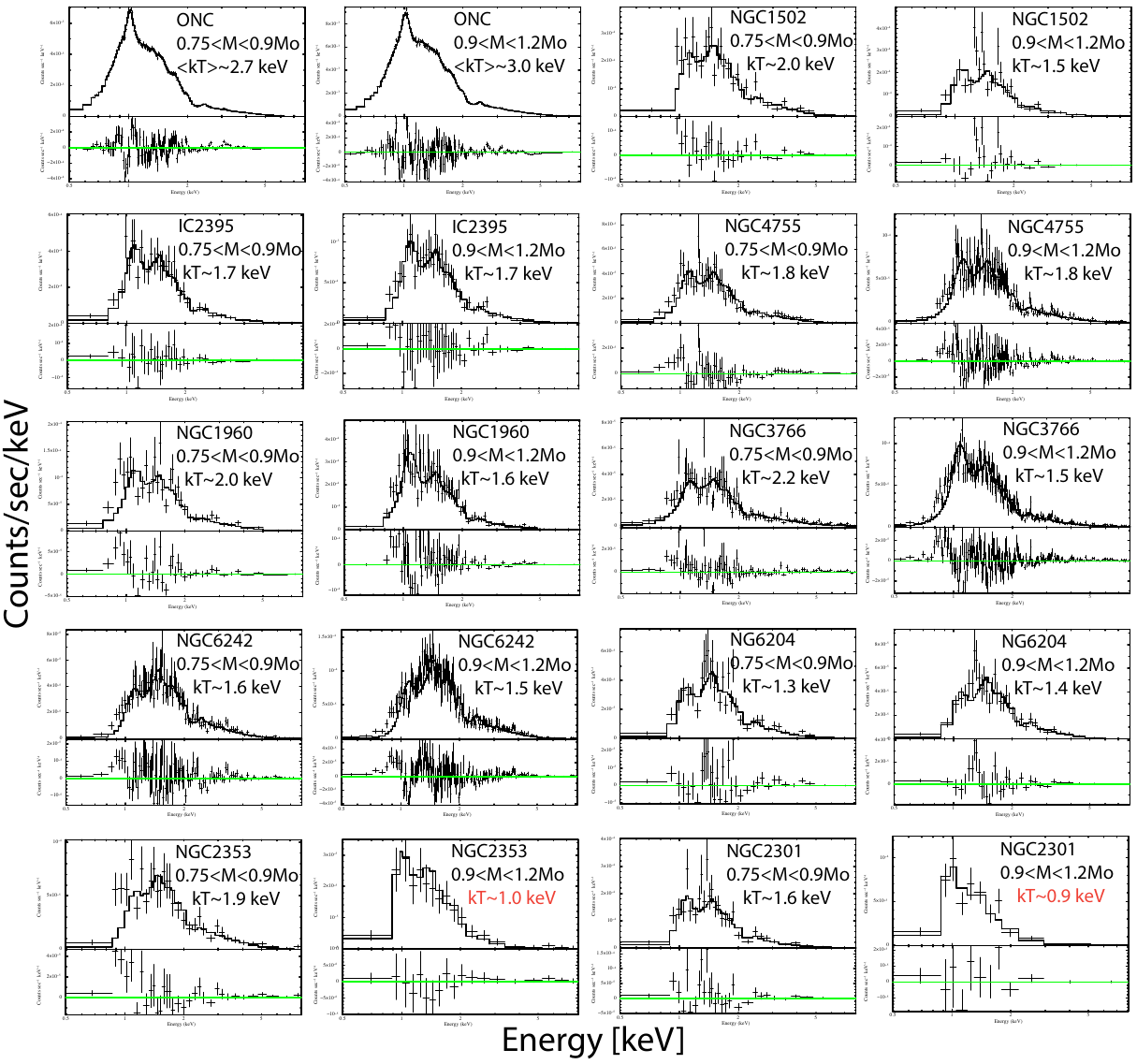}
\caption{Stacked {\it Chandra} spectra of solar-mass and sub-solar stellar members across clusters of different ages, overlaid with best-fit thermal plasma models. Each spectrum represents the combined X-ray emission from stars within a given mass and age group, fitted using XSPEC to derive average coronal temperatures. Stellar sample mass ranges and resulting average plasma temperatures are indicated in the figure legends. See Table~\ref{tab:xspec_fits_mass_strata} for details of the spectral fits.
\label{fig:xspec_spectra_two_mass_strata}}
\end{figure*}

\begin{deluxetable*}{ccccccccc}
%\tablenum{2}
\tabletypesize{\small}
\tablecaption{X-ray Spectral Fits of Stacked Data for Two Stellar Mass Strata in Various Clusters  \label{tab:xspec_fits_mass_strata}}
\tablewidth{0pt}
\tablehead{
\colhead{Cluster} & \colhead{Age} & \colhead{Mass Range} &  \colhead{$N_{stars}$} &
\colhead{$NC$} & \colhead{$\chi^{2}_\nu$} & \colhead{dof} & \colhead{$N_{H}$} & \colhead{$kT$} \\
\colhead{} & 
\colhead{(Myr)} & \colhead{($M_{\odot}$)} & \colhead{} & \colhead{(cnts)} & \colhead{} & \colhead{} & \colhead{$10^{22}$~cm$^{-2}$} & \colhead{(keV)}\\
\colhead{(1)} & \colhead{(2)} & \colhead{(3)} & \colhead{(4)} & \colhead{(5)} & \colhead{(6)} &
\colhead{(7)} & \colhead{(8)} &
\colhead{(9)}
}
%\decimalcolnumbers
\startdata
ONC & $<3$& $[0.75-0.9]$ & 33 & 169784 & 2.76 & 101 & 0.21 & $kT_1 = 0.95 \pm 0.02$ \\
\nodata & \nodata & \nodata & \nodata & \nodata & \nodata & \nodata & \nodata & $kT_2 = 5.03 \pm 0.25$ \\
\nodata & \nodata & \nodata & \nodata & \nodata & \nodata & \nodata & \nodata & $EM_2/EM_1 \sim 0.74$ \\
\nodata & \nodata & \nodata & \nodata & \nodata & \nodata & \nodata & \nodata & $kT_{av} = 2.70$  \\
ONC & $<3$& $[0.9-1.2]$ & 27 & 198901 & 2.76 & 113 & 0.23 & $kT_1 = 1.01 \pm 0.02$\\
\nodata & \nodata & \nodata & \nodata & \nodata & \nodata & \nodata & \nodata & $kT_2 = 5.49 \pm 0.21$ \\
\nodata & \nodata & \nodata & \nodata & \nodata & \nodata & \nodata & \nodata & $EM_2/EM_1 \sim 0.82$  \\
\nodata & \nodata & \nodata & \nodata & \nodata & \nodata & \nodata & \nodata & $kT_{av} = 3.02$  \\
NGC1502 & $\sim 7$  & $[0.75-0.9]$ & 20 & 489 & 1.41 & 29 & 0.42 & $1.98\pm0.20$ \\
NGC1502 & $\sim 7$ & $[0.9-1.2]$ & 24 & 441 & 2.16 & 25 & 0.42 & $1.48\pm0.22$ \\
NGC2169 & $\sim 7$ & $[0.75-0.9]$ & 4 & 124 & 1.44 & 6 & 0.14 & $1.27\pm0.17$ \\
NGC2169 & $\sim 7$ & $[0.9-1.2]$ & 9 & 300 & 1.24 & 17 & 0.14 & $1.49\pm0.11$ \\
IC2395 & $\sim 9$ & $[0.75-0.9]$ & 17 & 533 & 0.87 & 32 & 0.10 & $1.67\pm0.10$ \\
IC2395 & $\sim 9$ & $[0.9-1.2]$ & 9 & 664 & 0.86 & 38 & 0.10 & $1.68\pm0.09$ \\
NGC4755 & $\sim 15$ & $[0.75-0.9]$ & 121 & 875 & 1.20 & 55 & 0.24 & $1.78\pm0.14$ \\
NGC4755 & $\sim 15$ & $[0.9-1.2]$ & 175 & 2316 & 1.54 & 131 & 0.24 & $1.78\pm0.07$ \\
NGC1960 & $\sim 22$ & $[0.75-0.9]$ & 54 & 549 & 1.96 & 35 & 0.16 & $1.99\pm0.18$ \\
NGC1960 & $\sim 22$ & $[0.9-1.2]$ & 56 & 1201 & 1.45 & 65 & 0.16 & $1.56\pm0.06$ \\
NGC3766 & $\sim 22$ & $[0.75-0.9]$ & 121 & 1204 & 1.43 & 86 & 0.16 & $2.22\pm0.19$ \\
NGC3766 & $\sim 22$ & $[0.9-1.2]$ & 211 & 4008 & 1.43 & 178 & 0.16 & $1.50\pm0.03$ \\
NGC6242 & $\sim 45$ & $[0.75-0.9]$ & 90 & 1680 & 1.34 & 95 & 0.25 & $1.59\pm0.06$ \\
NGC6242 & $\sim 45$ & $[0.9-1.2]$ & 82 & 3168 & 1.63 & 151 & 0.25 & $1.49\pm0.03$ \\
NGC6204 & $\sim 75$ & $[0.75-0.9]$ & 39 & 520 & 1.38 & 34 & 0.28 & $1.33\pm0.09$ \\
NGC6204 & $\sim 75$ & $[0.9-1.2]$ & 34 & 518 & 1.02 & 33 & 0.28 & $1.43\pm0.09$ \\
NGC2353 & $\sim 100$ & $[0.75-0.9]$ & 15 & 458 & 1.66 & 31 & 0.08 & $1.91\pm0.27$ \\
NGC2353 & $\sim 100$ & $[0.9-1.2]$ & 25 & 220 & 0.86 & 18 & 0.08 & $0.97\pm0.09$ \\
NGC2301 & $\sim 100$ & $[0.75-0.9]$ & 19 & 445 & 1.60 & 28 & 0.05 & $1.58\pm0.12$ \\
NGC2301 & $\sim 100$ & $[0.9-1.2]$ & 24 & 158 & 0.76 & 10 & 0.05 & $0.90\pm0.09$ \\
\enddata
\tablecomments{Columns 1-2: Cluster name and approximate cluster age. Columns 3-4: Stellar mass range and number of stars included in the stack. Column 5: Total net X-ray counts in the stacked spectrum. Columns 6-7: Reduced $\chi^{2}$ for the overall spectral fit and degrees of freedom. Column 8: The X-ray absorption column density, in units of 10$^{22}$~cm$^{-2}$, was fixed during spectral fitting except for ONC. Column 9: Best-fit coronal plasma temperature(s) with 1~$\sigma$ uncertainties. For ONC, two-temperature fits ($tbabs \times (vapec + vapec)$) were performed; emission-measure-weighted average temperatures are also reported. Other clusters were fit with a single-temperature model ($tbabs \times apec$).}
\end{deluxetable*}

\subsection{Temporal evolution of coronal plasma  temperature}\label{sec:evolution_of_kt}

Stellar coronae are expected to cool and weaken with age as stellar rotation slows, reducing dynamo efficiency and magnetic reconnection activity \citep{Gudel2004}. For example, based on ROSAT and {\it ASCA} observations of nine solar-mass stars spanning a range of ages, and {\it Chandra} observations of $\sim 50$ low-mass stars in the 500 Myr-old M37 cluster, \citet{Gudel1997, Nunez2016} suggest that stars can lose their hot ($>1$--$3$ keV) coronal plasma component --- typically associated with large-scale magnetic structures (\S \ref{sec:discussion_magnetic_evolution}) --- by ages of $\sim 500$~Myr. Furthermore, the average coronal plasma temperature is expected to decrease with declining surface X-ray flux \citep{Johnstone2015}. Taking this into account, and given the observed steep decline in {\it  Chandra} X-ray luminosity among solar-mass stars by $\sim 100$~Myr (\S\ref{sec:results_temporal_evolution}), it is reasonable to investigate whether the average coronal plasma temperature in X-ray detected stars, as measured by {\it Chandra}, also exhibits significant changes across our large samples of solar- and lower-mass stars at different evolutionary stages. Our analysis is confined to the younger {\it Chandra} clusters reported by \citet{Getman2022}, together with five additional clusters aged 45--100 Myr recently observed with {\it Chandra} (Table~\ref{tab:cluster_props}).

The analyses presented here are conducted in two complementary ways:

\begin{enumerate}
\item A simpler approach based on evaluating the distributions of the apparent X-ray median energy ($ME$) for mass-stratified stellar members across clusters of different ages, without performing spectral fitting;

\item A more detailed method involving the stacking of mass-stratified X-ray spectra for clusters of different ages, followed by spectral fitting using XSPEC to derive coronal plasma temperatures.
\end{enumerate}

{\bf In the first experiment}, we use the {\it Acis Extract}-derived median energy of the net X-ray counts ($ME$) for each source of interest, as listed in Table~\ref{tab:xray_photometry} of the Appendix. $ME$ serves as a proxy for the coronal plasma temperature \citep{Getman10}, but it is an instrument-dependent quantity. Due to the gradual accumulation of contamination on {\it Chandra}'s optical blocking filter, the ACIS detector experiences temporal sensitivity degradation. As a result, $ME$ values for stellar samples observed at widely different epochs should be interpreted with caution and require systematic corrections. Such corrections for several epochs between 2003 and 2023 are calculated and presented in the Appendix of \citet{Getman2024}.

To avoid inter-epoch calibration issues, our $ME$-based analysis focuses not on comparing data across epochs, but rather on comparing $ME$ distributions (collected at similar epochs) among stellar samples of different masses but similar ages.

Figures~\ref{fig:me_evolution}(a-f) illustrate these comparisons, showing $ME$ distributions for solar-mass (red), sub-solar-mass (green), and lower-mass (grey) stars, grouped by similar cluster ages and observation epochs. To increase the sample size for statistical analysis, stellar samples from clusters of similar ages and observed at close epochs are combined. Our primary scientific focus is on the $ME$ values for the X-ray detected solar-mass ($0.9$--$1.2$~M$_{\odot}$) and sub-solar-mass ($0.75$--$0.9$~M$_{\odot}$) stars, for which the mass completeness is ensured. The lower-mass stars ($<0.75$~M$_{\odot}$;  grey), which are not mass-complete, are included for illustrative purposes only.

The main result from this first, $ME$-based, experiment is that a statistically significant difference in coronal plasma temperature between X-ray  detected solar- and sub-solar-mass stars emerges only around an age of $\sim 100$~Myr. At this stage, solar-mass stars exhibit cooler coronal temperatures than sub-solar-mass stars, with high statistical significance (Anderson-Darling test, $p$-value $< 0.001$).

In the case of the older $\sim 100$~Myr NGC~2353 and NGC~2301 clusters, we also used XSPEC's {\it fakeit} command in conjunction with averaged {\it Chandra} ARFs produced by {\it ACIS Extract} to simulate a grid of one-temperature {\it apec} spectra (subject to typical cluster' low absorption via {\it tbabs}) spanning a wide range of coronal plasma temperatures. Separately, we employed XSPEC's {\it iplot} and {\it wecum} commands to generate cumulative distribution functions (CDFs) of the observed X-ray net counts for the solar-mass and sub-solar-mass stellar samples in these clusters. Consistent with the earlier $ME$-based results, Figure~\ref{fig:me_evolution}(g) overlays the observed CDFs on the simulated expectations and shows that the solar-mass stars exhibit $>80$\% of net counts in line with an average plasma temperature of $\sim 1$~keV. In contrast, the sub-solar sample requires much hotter average temperatures of $\sim 1.6–1.9$~keV. 

Importantly, despite the declining soft-energy sensitivity of {\it Chandra}-ACIS-I, the instrument remains capable of detecting cooler young stars --- with average plasma temperatures as low as $kT \sim 0.4$~keV --- provided they exhibit X-ray brightness comparable to the solar-mass detections in NGC~2353 and NGC~2301. If such stars were present, the empirical {\it Chandra} CDF of X-ray emission from dominant $kT \sim 0.4$~keV coronal plasma would reach the 85\% level at 1.5~keV, compared to the current 60\%. 

{\bf In the second experiment}, we follow the same methodology as described in \S\ref{sec:Xray_luminosities}. Specifically, we stack the {\it Chandra} spectra of solar-mass and sub-solar stars separately for each cluster, and fit the resulting spectra in XSPEC to derive average plasma temperatures. As in the previous analysis, column densities and coronal abundances are held fixed.

Due to limited photon statistics in most samples (except the ONC), we adopt a simplified one-temperature spectral model, $tbabs \times apec$, with free plasma temperature ($kT$) and emission measure ($EM=10^{14} \times norm \times 4 \pi D^2$) parameters. For the ONC, where much deeper {\it Chandra} data from the long-duration COUP project are available \citep{Getman05,Getman2024}, we apply a more complex two-temperature model, $tbabs \times (vapec + vapec)$. For the ONC, the resulting cool and hot temperature components are then averaged, weighted by their respective emission measures. The corresponding spectral fit results are summarized in Table~\ref{tab:xspec_fits_mass_strata}, and the fitted spectra are shown in Figure~\ref{fig:xspec_spectra_two_mass_strata}.

In principle, the correction factors ($CF_{AE}$) derived from these more mass-specific subsamples could have been applied in the $PFlux$--$L_X$ conversion described in \S~\ref{sec:Xray_luminosities}. However, they are statistically indistinguishable from the $CF_{AE}$ values obtained using the full X-ray member samples listed in Table~\ref{tab:xspec_fits}. For example, even for the oldest and X-ray--softest $0.9$--$1.2$~M$_{\odot}$ subsamples in NGC~2353 and NGC~2301, the fits in Table~\ref{tab:xspec_fits_mass_strata} yield $CF_{AE} = 3.10 \pm 0.79$ and $3.31 \pm 0.80$ keV per photon, respectively. These values differ by less than $1.3\sigma$ from the corresponding correction factors for the full X-ray samples --- $4.23 \pm 0.25$ and $2.79 \pm 0.18$ keV per photon (Table~\ref{tab:xspec_fits}) --- and are therefore statistically consistent. 

This experiment reinforces the results from the $ME$-based analysis: the stacked spectra of solar-mass stars in the $\sim 100$~Myr-old clusters NGC~2353 and NGC~2301 require significantly softer coronal emission, with $kT \leq 1$~keV. In contrast, spectra of younger solar-mass stars and all sub-solar stars in ($1$--$100$)~Myr clusters exhibit hotter plasma components in the $1.3$--$3$~keV range. This suggests that solar-mass stars begin to lose their hot ($>$1--$3$~keV) coronal plasma --- associated with large-scale surface and volume structures (\S\ref{sec:discussion_magnetic_evolution})  --- by an age of $\sim 100$~Myr, indicating a more rapid temperature decline than previously assumed \citep{Gudel1997}.

\section{Discussion} \label{sec:discussion}
%% LEAVE THIS STATEMENT FOR DISCUSSION 
%%Interestingly, by the age of 500 Myr—when most low-mass stars, except for the very lowest-mass ones ($<0.3$M$_\odot$), have already reached the ZAMS—their hot coronal plasma component ($kT_2 > 1$–2keV) appears to vanish. For example, \citet{Nunez2016} find that only a soft plasma component ($kT_1 \sim 0.9$~keV) is required to fit the {\it Chandra} X-ray spectra of numerous low-mass members of the M37 open cluster. STILL overwrite becauses Nunez is likely not sensitive to $<$0.5Mo stars

\subsection{Magnetic Dynamo Evolution and Coronal Transition in Solar-Mass Stars} \label{sec:discussion_magnetic_evolution}

Our X-ray-Gaia study of open clusters spanning 7--750~Myr reveals a pronounced contrast in the temporal evolution of X-ray activity between solar-mass stars (0.9--1.2~M$_{\odot}$) and slightly lower-mass stars (0.75--0.9~M$_{\odot}$). Both groups exhibit declining X-ray luminosities with age, but during the first hundred Myr, solar-mass stars experience a significantly steeper decay ($L_X \propto t^{-1.1}$) compared to the shallower trend ($L_X \propto t^{-0.5}$) of their lower-mass counterparts (\S \ref{sec:results_temporal_evolution}, Figure~\ref{fig:lx_vs_t_main}). An alternative view of Figure~\ref{fig:lx_vs_t_main} suggests that solar-mass stars may track the slower decline until $\sim$45~Myr, followed by a sharp drop between 45 and 100~Myr. 

Concurrently, solar-mass stars near $\sim 100$~Myr show a marked softening of their X-ray spectra, with a substantial reduction of the hot coronal plasma component ($kT \gtrsim 1$--3~keV; \S\ref{sec:evolution_of_kt}). These empirical trends indicate a key transition in the operation of magnetic dynamos, coronal heating, and overall magnetic activity in young solar-type stars.

Furthermore, these results build upon and extend earlier findings by \citet{Getman2022,Getman2023,Getman2025} and others, which demonstrated that young stars across a broad mass range exhibit a nearly universal relation between X-ray luminosity and surface magnetic flux, $L_X \propto \Phi^1$, consistent with that observed in main-sequence stars and the Sun \citep{Pevtsov2003,Toriumi2022}. For solar-mass stars younger than 25 Myr, both X-ray luminosity and surface spot area decline with time, following $L_X \propto t^{-1}$ and $A_{\rm spot} \propto t^{-1}$. This behavior is consistent with a scenario in which the magnetic field strength in starspots ($B_{\rm spot} \sim 3$~kG) remains approximately constant, while the spot filling factor ($f_{\rm spot}$) decreases with age \citep{Kochukhov2020}. 

By $\sim$40--50~Myr, solar-mass stars complete their contraction onto the zero-age main sequence (as indicated by PARSEC models) and begin to experience rotational spin-down, driven by efficient angular momentum loss via magnetized stellar winds \citep{Amard2019,Gossage2021}. As these stars spin down and their convective envelopes thin with age, the properties of their magnetic dynamos are expected to evolve significantly.

Theoretical studies suggest that the efficiency and character of stellar dynamos are tightly linked to both rotation and convection zone structure. Fully convective stars support efficient $\alpha^2$ distributed dynamos \citep{Browning2008}, while solar-type stars with radiative cores rely on $\alpha \Omega$ interface (tachocline) dynamos operating at the boundary between convective and radiative zones \citep{Brun2004,Charbonneau2020,Kapyla2023}. The effectiveness of these interface dynamos depends on both the depth of the convection zone and the amplitude of differential rotation across the tachocline \citep{Jouve2010,Brown2011}. Interface dynamo theory predicts that dynamo efficiency scales inversely with the square of the Rossby number, which incorporates both the stellar rotation period and convective overturn timescale as key parameters \citep{Montesinos2001}. Evolutionary models and observational studies show that as solar-mass stars age, their convective turnover times decrease while rotation periods increase, causing the Rossby number to rise \citep{Landin2010,Wright2011,Landin2023}. This increasing Rossby number drives the transition from saturated to unsaturated magnetic regimes and correlates with the observed decline in magnetic activity.

The decreasing dynamo efficiency in solar-mass stars leads to smaller maximum surface spot areas and reduced coronal heating rates, resulting in lower coronal temperatures and smaller pressure scale heights of X-ray-emitting coronal structures. The X-ray luminosity is given by $L_X = V n_e^2 P(T)$, where $V$ is the X-ray emitting volume, $n_e$ is the electron density, and $P(T)$ is the temperature-dependent radiative loss function for optically thin coronal plasma \citep{Rosner1978,Sutherland1993}. Assuming approximately constant $n_e$, and using observed values of $L_X$ and $kT$ for solar-mass stars at different evolutionary stages (see Tables~\ref{tab:temporal_evolution_table}, \ref{tab:xspec_fits_mass_strata}), along with the empirical correlation $L_X \propto f_{\rm spot}^{2.7}$ \citep{Kochukhov2020}, one can infer that the characteristic height of X-ray coronal structures decreases by roughly a factor of two between 40 and 100 Myr. 

An independent comparison of lower-mass stars in the $1$--$2$~Myr-old ONC and the $\sim 100$~Myr-old Pleiades \citep{Getman08b,Guarcello2019} shows that X-ray flaring coronal structures shrink over time, from $1$--$10$~R$_{\star}$ to $<0.5$--$1$~R$_{\star}$. 

Larger coronal structures in younger stars and in the Sun are likely powered by magnetic reconnection processes, which dominate over less efficient Alfv\'{e}n wave heating mechanisms \citep{Toriumi2022, Getman2025}, and are typically associated with higher plasma temperatures \citep{Aschwanden2008,Getman2025}.

The disappearance of the hot coronal component ($kT > 1$--3~keV) in solar-mass stars near 100 Myr is therefore likely a direct observational manifestation of declining dynamo efficiency. The corona becomes dominated by smaller, lower-altitude loops filled with cooler plasma, similar to those seen in the contemporary Sun. In contrast, lower-mass stars (0.75--0.9~M$_\odot$) retain deeper convective envelopes and higher rotation rates at this age, likely sustaining more efficient convection-driven dynamos and preserving their hotter and larger coronal structures.

We therefore propose that magnetic activity in solar-mass stars undergoes a relatively rapid transition between $\sim$40 and 100 Myr, driven by changes in stellar structure (envelope thinning), rotation (spin-down), and magnetic dynamo regime (from distributed to interface-type). This transition is reflected in the observed steep decline of X-ray luminosity and the disappearance of the hot coronal plasma component in our stellar sample.

\subsection{Possible Implications for  Planetary Environments at $\sim 100$--750~Myr} \label{sec:discussion_implications}

Calibrated to empirical $(0.5$--$7)$~keV X-ray luminosities from \citet{Nunez2016}, the semi-analytic activity-rotation model of \citet{Johnstone2021} predicts an extended plateau in X-ray activity for solar-mass stars between $\sim 40$ and 300~Myr, with $\log L_X$ remaining near $\sim 29.5$~erg~s$^{-1}$ (Figure~\ref{fig:lx_vs_t_main}). Notably, this model is widely used by the planetary science community to assess the impact of stellar XUV irradiation on exoplanet atmospheres.

Our {\it Chandra}$+$Gaia analysis of numerous $\lesssim 100$~Myr clusters, combined with archival {\it ROSAT} and {\it Chandra} data and Gaia-based membership for older clusters (NGC~6475, M37, Praesepe), demonstrates that solar-mass stars experience a stronger and more continuous decline in X-ray activity than previously assumed. Rather than an extended activity plateau between 40--300~Myr, we observe a monotonic dimming: from $\log L_X \sim 29.5$~erg~s$^{-1}$ at $\sim 45$~Myr, to $\log L_X \sim 29$~erg~s$^{-1}$ at $\sim 100$~Myr, and to below $\sim 28.5$~erg~s$^{-1}$ by $\sim 750$~Myr, accompanied by systematic coronal softening. This offset of $\gtrsim 0.5$~dex relative to common model prescriptions carries significant implications for planetary atmospheric evolution near 100~Myr and continuing to later epochs up to $\sim 750$~Myr:

\begin{itemize} 

\item {\bf Atmospheric Mass Loss of Young Planets.}  
Primordial H/He envelopes around close-in Earth-mass planets are expected to be rapidly stripped, often within $\lesssim 10$ Myr, under the intense XUV irradiation of their young host stars. This process, modeled in photoevaporation frameworks \citep{Owen2017} and supported by empirical X-ray measurements of young solar analogs \citep{Getman2021,Getman2022}, helps explain the absence of extended atmospheres on many close-in terrestrial planets. By contrast, sub-Neptune and mini-Neptune planets with more massive H/He envelopes experience their most vigorous hydrodynamic mass loss over the subsequent few hundred Myr, while stellar XUV output remains elevated \citep{Owen2019}. However, models that assume a high and sustained activity plateau in solar-type stars through $\sim 300$ Myr, informed by the XUV results of \citet{Johnstone2021}, likely overestimate the extent of such atmospheric erosion. 

Recent theoretical work highlights the sensitivity of escape efficiency to the stellar XUV flux and planetary density \citep{McCreery2025}, the role of photoevaporation in volatile partitioning during the first few hundred Myr \citep{Tomberg2024}, and the transitions between core-powered and photoevaporative escape depending on the penetration depth of stellar photons \citep{Owen2024, Rogers2024}. Other studies demonstrate that stellar XUV evolution directly shapes the location and slope of the radius valley and the related ``radius cliff'' \citep{Dattilo2024, King2024, Affolter2023}. Observations also confirm that XUV-driven erosion can transform mini-Neptunes into super-Earths on Gyr timescales \citep{Malsky2023, Zhang2023}. Observational case studies of young planetary systems likewise highlight the importance of high-energy environments in driving atmospheric erosion, even for relatively inactive hosts \citep{Poppenhaeger2024, Ketzer2024}. Within this context, our lower empirical X-ray activity track suggests that many young planets may retain larger fractions of their primordial H/He envelopes than previously estimated, warranting revisions to modern photoevaporative escape models as well as a reassessment of radius valley demographics.

\item {\bf Habitability Windows and Water Retention.}  
The reduced X-ray flux and softer coronal spectra at 100--750~Myr also mitigate photolytic loss of hydrogen from H$_2$O and other volatile species. Compared to prior expectations, terrestrial planets in the habitable zones of G-type stars may experience slower escape of hydrogen generated by water photolysis, and a decreased likelihood of accumulating large abiotic O$_2$ reservoirs \citep{Wordsworth2014, Harman2015}. Hydrodynamic simulations show that volatile retention between 50--500 Myr is strongly modulated by stellar XUV flux and winds \citep{Canet2024}. The moderating effect in our revised stellar activity track therefore extends the potential window for water retention into mid-Gyr epochs, allowing nascent terrestrial planets to remain water-rich with relatively thick prebiotic atmospheres, and reduces the probability of false-positive oxygen biosignatures. By contrast, K-type stars, which show a slower decay of X-ray activity, may maintain harsher irradiation conditions for longer, thereby imposing stronger constraints on volatile retention and habitability outcomes. 
  
\item {\bf Prebiotic Chemistry and Atmospheric Ionization.}  
High-energy radiation also influences the production of prebiotic molecules and the evolution of atmospheric compositions. Recent models, such as those by \citet{Locci2022}, demonstrate how X-rays penetrate deeply into planetary atmospheres, producing ionization cascades that lead to synthesis of complex organics. Similarly, \citet{Barth2021} show that XUV-driven chemistry, including energetic particles, enhances the formation of prebiotic feedstock molecules such as HCN, CH$_2$O, and C$_2$H$_4$. Additional work shows that XUV-driven compositional fractionation can generate vertical atmospheric gradients \citep{Modirrousta-Galian2024}, further complicating assessments of whether stellar irradiation fosters or hinders prebiotic chemistry. The net prebiotic impact of stellar irradiation thus depends on a delicate balance: low-to-moderate ionizing fluxes may promote chemical synthesis in upper atmospheres or hydrocarbon hazes, while extreme flare events or prolonged XUV exposure may destroy molecular precursors or trigger complete atmospheric erosion \citep{Airapetian2016,Yamashiki2019,Barth2021}. Our findings suggest that solar-type stars between 100--750~Myr may generally provide lower ionizing fluxes than assumed, potentially shifting the balance toward more favorable conditions for prebiotic chemical synthesis.  

\end{itemize}

\section{Conclusions} \label{sec:conclusions}

This study extends the X-ray activity-mass-age analysis of stars younger than 25 Myr by \citet{Getman2022} to a broader age range, reaching up to $\sim 750$~Myr. We achieve this by analyzing new {\it Chandra} observations of five additional rich open clusters aged $\sim 45$--100~Myr (\S \ref{sec:targets_chandra}), incorporating published {\it ROSAT} and {\it Chandra} measurements for three older clusters ($\sim 220$--750~Myr; \S \ref{sec:3more_older_clusters}), performing joint X-ray-Gaia analyses to refine cluster distances, ages, memberships, and stellar properties (\S\S \ref{sec_membership}--\ref{sec:3more_older_clusters}), and evaluating mass-stratified trends in stellar X-ray luminosity and coronal temperature as functions of age (\S \ref{sec:resuts}).

Our key results reveal a mass-dependent decay in X-ray luminosity with age, with solar-mass stars exhibiting a significantly stronger decline and softening of coronal emission compared to their lower-mass siblings. These trends are likely linked to a reduction in magnetic dynamo efficiency and a diminished ability to sustain large-scale, high-temperature coronal structures in solar-mass stars (\S \ref{sec:discussion_magnetic_evolution}).

This stronger-than-expected dimming and softening of X-ray emission in solar-mass stars from $\sim 100$ to $\sim 750$~Myr, revealed through comparison of our empirical results with the widely used semi-analytic activity-rotation model of \citet{Johnstone2021}, carries important implications for the assessment of planetary evolution. It suggests reduced rates of atmospheric mass loss and water photolysis, along with changes in ionization levels and chemical pathways relevant to prebiotic molecule formation (\S\ref{sec:discussion_implications}).

\begin{acknowledgments}
 We are grateful to the anonymous referee for providing thoughtful and helpful comments that improved the manuscript. We thank Patrick Broos (Penn State) for his help with running the {\it Acis Extract} software. This project is supported by the {\it Chandra} grant GO3-24007X (K. Getman, Principal Investigator) and the {\it Chandra} ACIS Team contract SV474018 (G. Garmire \& L. Townsley, Principal Investigators), issued by the {\it Chandra} X-ray Center, which is operated by the Smithsonian Astrophysical Observatory for and on behalf of NASA under contract NAS8-03060. The {\it Chandra} Guaranteed Time Observations (GTO) data used here were selected by the ACIS Instrument Principal Investigator, Gordon P. Garmire, of the Huntingdon Institute for X-ray Astronomy, LLC, which is under contract to the Smithsonian Astrophysical Observatory; contract SV2-82024. V. S. A. was supported by the GSFC Sellers Exoplanet Environments Collaboration (SEEC), which is funded by the NASA Planetary Science Divisions Internal Scientist Funding Model (ISFM), the NASA NNH21ZDA001N-XRP F.3 Exoplanets Research Program grants. The Center for Exoplanets and Habitable Worlds is supported by the Pennsylvania State University and the Eberly College of Science. This work has made use of data from the European Space Agency (ESA) mission {\it Gaia} (\url{https://www.cosmos.esa.int/gaia}), processed by the {\it Gaia} Data Processing and Analysis Consortium (DPAC, \url{https://www.cosmos.esa.int/web/gaia/dpac/consortium}). Funding for the DPAC has been provided by national institutions, in particular the institutions participating in the {\it Gaia} Multilateral Agreement. This paper employs a list of Chandra datasets, obtained by the Chandra X-ray Observatory, contained in~\dataset[DOI: 10.25574/cdc.515]{https://doi.org/10.25574/cdc.515}.
\end{acknowledgments}

%%\begin{contribution}
%%This section gives authors the space to recognize author contributions. The text inside this environment is NOT counted towards the total word quanta. At a minimum, manuscripts are expected to include this text:

%%All authors contributed equally to the Terra Mater collaboration.

%% But authors are expected to provide more specific details, e.g. 
%%
%%SC was responsible for writing and submitting the manuscript.
%%WWM came up with the initial research concept and edited the manuscript.
%%OTS obtained the funding and edited the manuscript.
%%EBF provided the formal analysis and validation. He also edited the manuscript.
%%GEH Supervised the undergraduates, wrote the software and administers the project github and Zenodo repositories.
%%
%% Authors can use the Contributor Role Taxonomy (CRediT) at
%% https://credit.niso.org
%% for ideas on how write a good statement tailored to their needs.

%%\end{contribution}

%% To help institutions obtain information on the effectiveness of their 
%% telescopes the AAS Journals has created a group of keywords for telescope 
%% facilities.
%
%% Following the acknowledgments section, use the following syntax and the
%% \facility{} or \facilities{} macros to list the keywords of facilities used 
%% in the research for the paper.  Each keyword is check against the master 
%% list during copy editing.  Individual instruments can be provided in 
%% parentheses, after the keyword, but they are not verified.
\facilities{CXO, Gaia, ROSAT}

%% Similar to \facility{}, there is the optional \software command to allow 
%% authors a place to specify which programs were used during the creation of 
%% the manuscript. Authors should list each code and include either a
%% citation or url to the code inside ()s when available.
\software{R \citep{RCoreTeam20}, Acis Extract \citep{Broos10,Broos2012}, CIAO \citep{Fruscione2006}, XSPEC \citep{Arnaud1996}}

%% Appendix material should be preceded with a single \appendix command.
%% There should be a \section command for each appendix. Mark appendix
%% subsections with the same markup you use in the main body of the paper.
%%
%% Each Appendix (indicated with \section) will be lettered A, B, C, etc.
%% The equation counter will reset when it encounters the \appendix
%% command and will number appendix equations (A1), (A2), etc. The
%% Figure and Table counter will not reset.

\appendix

\section{Supplementary Material} \label{sec:appendix}

This section contains the following supplementary materials: a table listing the {\it Chandra} observations of the five $\sim 45$--100~Myr open clusters (Table~\ref{tab:log_chandra_observations}); a table listing the X-ray properties of all detected sources across the {\it Chandra} fields of these five $\sim 45$--100~Myr open clusters (Table~\ref{tab:xray_photometry}); figures summarizing member identification and characterization for four of the five $\sim 45$--100~Myr open clusters (Figures~\ref{fig:mem_NGC6204}--\ref{fig:mem_NGC2301}); and a figure showing the inferred temporal evolution of X-ray luminosity across six additional stellar-mass strata (Figure~\ref{fig:lx_vs_t_appendix}).

\begin{deluxetable}{lrccccr}
%\tablenum{2}
\tabletypesize{\tiny}
\tablecaption{Log of {\it Chandra}-ACIS-I Observations  \label{tab:log_chandra_observations}}
\tablewidth{0pt}
\tablehead{
\colhead{Cluster} & \colhead{ObsId} &
\colhead{Exposure} & \colhead{R.A.} & \colhead{Decl.} & \colhead{Start Time} & \colhead{PI}\\
\colhead{} & \colhead{} &
\colhead{(ksec)} & \colhead{$\alpha_{J2000}$} & \colhead{$\delta_{J2000}$} & \colhead{(UT)} & \colhead{}\\
\colhead{(1)} & \colhead{(2)} & \colhead{(3)} & \colhead{(4)} & \colhead{(5)} & \colhead{(6)} &
\colhead{(7)}
}
%\decimalcolnumbers
\startdata
NGC 6242 & 27043 & 11.9 & 16:55:31.76 & -39:28:05.5 & 2024-01-27 & K. Getman \\
NGC 6242 & 27041 & 10.0 & 16:55:32.13 & -39:28:09.3 & 2024-03-10 & K. Getman \\
NGC 6242 & 27040 & 27.6 & 16:55:32.04 & -39:28:08.7 & 2024-03-19 & K. Getman \\
NGC 6242 & 29335 & 29.7 & 16:55:32.13 & -39:28:11.7 & 2024-03-21 & K. Getman \\
NGC 6242 & 27037 & 23.8 & 16:55:32.40 & -39:28:12.3 & 2024-03-22 & K. Getman \\
NGC 6242 & 26499 & 18.3 & 16:55:32.67 & -39:28:17.0 & 2024-05-09 & K. Getman \\
NGC 6242 & 29213 & 10.4 & 16:55:32.73 & -39:28:24.3 & 2024-05-15 & K. Getman \\
NGC 6242 & 27039 & 16.4 & 16:55:32.84 & -39:28:26.4 & 2024-05-16 & K. Getman \\
NGC 6242 & 29408 & 15.5 & 16:55:32.69 & -39:28:26.0 & 2024-05-19 & K. Getman \\
NGC 6242 & 27038 & 11.2 & 16:55:31.95 & -39:28:44.8 & 2024-06-07 & K. Getman \\
NGC 6242 & 29449 & 19.8 & 16:55:31.60 & -39:28:48.6 & 2024-06-11 & K. Getman \\
NGC 6242 & 27042 & 32.1 & 16:55:29.75 & -39:28:45.5 & 2024-08-16 & K. Getman \\
NGC 6242 & 29214 & 10.0 & 16:55:29.58 & -39:28:45.2 & 2024-08-26 & K. Getman \\
NGC 6242 & 27036 & 19.7 & 16:55:29.14 & -39:28:39.2 & 2024-09-13 & K. Getman \\
NGC 6242 & 29441 & 10.5 & 16:55:29.09 & -39:28:43.2 & 2024-09-20 & K. Getman \\
NGC 6242 & 29442 & 10.0 & 16:55:29.42 & -39:28:45.2 & 2024-09-22 & K. Getman \\
Trumpler 3 & 27414 & 14.7 & 03:11:57.11 & +63:13:06.3 & 2022-12-04 & G. Garmire \\
Trumpler 3 & 27587 & 12.9 & 03:11:57.13 & +63:13:06.4 & 2022-12-04 & G. Garmire \\
Trumpler 3 & 27413 & 19.8 & 03:11:57.56 & +63:13:11.7 & 2023-11-23 & G. Garmire \\
Trumpler 3 & 29078 & 13.6 & 03:11:57.24 & +63:13:10.3 & 2023-11-24 & G. Garmire \\
Trumpler 3 & 27348 & 35.8 & 03:11:58.57 & +63:12:51.1 & 2024-01-13 & G. Garmire \\
Trumpler 3 & 28830 & 27.0 & 03:12:00.76 & +63:13:26.0 & 2024-10-02 & G. Garmire \\
Trumpler 3 & 28829 & 10.0 & 03:11:59.87 & +63:12:41.2 & 2025-02-11 & G. Garmire \\
Trumpler 3 & 30791 & 10.0 & 03:11:59.83 & +63:12:41.2 & 2025-02-14 & G. Garmire \\
Trumpler 3 & 30792 & 12.9 & 03:11:59.92 & +63:12:41.7 & 2025-02-15 & G. Garmire \\
Trumpler 3 & 28762 & 10.0 & 03:12:02.06 & +63:12:43.6 & 2025-04-18 & G. Garmire \\
Trumpler 3 & 30897 & 10.0 & 03:12:02.06 & +63:12:43.8 & 2025-04-19 & G. Garmire \\
Trumpler 3 & 30898 & 10.0 & 03:12:02.00 & +63:12:43.9 & 2025-04-19 & G. Garmire \\
NGC 6204 & 26498 & 13.7 & 16:46:10.33 & -47:01:19.1 & 2023-01-20 & K. Getman \\
NGC 6204 & 27668 & 13.7 & 16:46:10.28 & -47:01:18.0 & 2023-01-20 & K. Getman \\
NGC 6204 & 27243 & 21.8 & 16:46:10.69 & -47:01:20.9 & 2024-03-11 & K. Getman \\
NGC 6204 & 29329 & 21.8 & 16:46:10.71 & -47:01:20.6 & 2024-03-11 & K. Getman \\
NGC 6204 & 27241 & 21.3 & 16:46:10.64 & -47:01:20.0 & 2024-03-11 & K. Getman \\
NGC 6204 & 27242 & 24.8 & 16:46:11.11 & -47:01:26.6 & 2024-04-03 & K. Getman \\
NGC 6204 & 27240 & 14.3 & 16:46:11.20 & -47:01:44.3 & 2024-05-27 & K. Getman \\
NGC 6204 & 27245 & 16.7 & 16:46:11.28 & -47:01:46.9 & 2024-05-30 & K. Getman \\
NGC 6204 & 29425 & 16.3 & 16:46:11.20 & -47:01:47.1 & 2024-05-30 & K. Getman \\
NGC 6204 & 29426 & 10.0 & 16:46:11.09 & -47:01:49.9 & 2024-06-01 & K. Getman \\
NGC 6204 & 27246 & 32.9 & 16:46:07.48 & -47:01:53.7 & 2024-09-11 & K. Getman \\
NGC 6204 & 27244 & 12.7 & 16:46:07.43 & -47:01:53.0 & 2024-09-22 & K. Getman \\
NGC 6204 & 30079 & 16.9 & 16:46:07.50 & -47:01:52.9 & 2024-09-22 & K. Getman \\
NGC 2353 & 26500 & 42.2 & 07:14:35.19 & -10:15:10.7 & 2022-11-24 & K. Getman \\
NGC 2353 & 27044 & 15.6 & 07:14:34.27 & -10:15:00.0 & 2023-08-31 & K. Getman \\
NGC 2353 & 27046 & 9.8 & 07:14:34.90 & -10:15:08.7 & 2023-10-25 & K. Getman \\
NGC 2353 & 29003 & 9.8 & 07:14:34.94 & -10:15:11.6 & 2023-10-27 & K. Getman \\
NGC 2353 & 27416 & 14.9 & 07:14:34.87 & -10:15:07.1 & 2023-11-02 & G. Garmire \\
NGC 2353 & 29030 & 14.9 & 07:14:35.09 & -10:15:10.2 & 2023-11-04 & G. Garmire \\
NGC 2353 & 27048 & 24.8 & 07:14:35.13 & -10:15:11.4 & 2023-11-10 & K. Getman \\
NGC 2353 & 29053 & 15.9 & 07:14:35.20 & -10:15:11.1 & 2023-11-10 & G. Garmire \\
NGC 2353 & 27049 & 17.7 & 07:14:35.30 & -10:15:15.0 & 2023-11-30 & K. Getman \\
NGC 2353 & 29091 & 19.4 & 07:14:35.30 & -10:15:13.7 & 2023-12-02 & K. Getman \\
NGC 2353 & 27349 & 7.6 & 07:14:33.98 & -10:15:49.0 & 2024-02-04 & G. Garmire \\
NGC 2353 & 27415 & 13.7 & 07:14:33.89 & -10:15:48.1 & 2024-02-18 & G. Garmire \\
NGC 2353 & 27045 & 42.3 & 07:14:33.14 & -10:15:44.1 & 2024-03-12 & K. Getman \\
NGC 2353 & 29042 & 14.9 & 07:14:32.66 & -10:15:40.5 & 2024-04-16 & K. Getman \\
NGC 2353 & 27047 & 14.2 & 07:14:32.63 & -10:15:38.9 & 2024-04-19 & K. Getman \\
NGC 2353 & 29224 & 14.1 & 07:14:34.33 & -10:15:02.1 & 2024-09-07 & G. Garmire \\
NGC 2301 & 26002 & 17.8 & 06:51:47.17 & +00:27:54.7 & 2021-12-17 & G. Garmire \\
NGC 2301 & 26238 & 19.8 & 06:51:47.16 & +00:27:55.1 & 2021-12-19 & G. Garmire \\
NGC 2301 & 26082 & 15.8 & 06:51:47.04 & +00:27:49.2 & 2022-01-03 & G. Garmire \\
NGC 2301 & 26261 & 13.9 & 06:51:46.93 & +00:27:45.2 & 2022-01-06 & G. Garmire \\
NGC 2301 & 26081 & 15.7 & 06:51:45.59 & +00:27:48.8 & 2022-05-14 & G. Garmire \\
NGC 2301 & 26418 & 14.9 & 06:51:45.59 & +00:27:47.4 & 2022-05-14 & G. Garmire \\  
\enddata
\tablecomments{Column 1: Cluster name. Columns 2-3: {\it Chandra}-ACIS-I observation identifier (ObsID) and net exposure time in kiloseconds. Columns 4-5:  Aimpoint of the {\it Chandra} observation in right ascension and declination.  Column 6: Start calendar day of {\it Chandra} exposure. Column 7: Principal investigator of {\it Chandra} observation.}
\end{deluxetable}

\begin{deluxetable*}{ccccrcccc}
%\tablenum{2}
\tabletypesize{\small}
\tablecaption{X-ray Sources Towards 5 New Open Clusters \label{tab:xray_photometry}}
\tablewidth{0pt}
\tablehead{
\colhead{Cluster} & \colhead{CXOU J} & \colhead{R.A.} &
\colhead{Decl.} & $C_{net}$& \colhead{$\sigma_{net}$} & \colhead{$PFlux$} & \colhead{$ME$} &
\colhead{Group} \\
\colhead{} & \colhead{} &  \colhead{(deg)} &
\colhead{(deg)} & \colhead{(cnts)} & \colhead{(cnts)} & \colhead{(ph cm$^{-2}$ s$^{-1}$)} & \colhead{(keV)} &
\colhead{}\\
\colhead{(1)} & \colhead{(2)} & \colhead{(3)} & \colhead{(4)} & \colhead{(5)} &
\colhead{(6)} & \colhead{(7)} & \colhead{(8)} & \colhead{(9)} 
}
%\decimalcolnumbers
\startdata
NGC 6242 & 165511.96-393120.3 & 253.799842 & -39.522316 & 46.5 & 7.0 & -5.995 & 1.9 & 5\\
NGC 6242 & 165511.99-392407.3 & 253.799983 & -39.402038 & 13.9 & 4.4 & -6.501 & 1.5 & 6\\
NGC 6242 & 165511.99-393209.1 & 253.799994 & -39.535884 & 39.2 & 6.6 & -6.068 & 2.6 & 6\\
NGC 6242 & 165512.00-392302.1 & 253.800010 & -39.383923 & 12.4 & 4.5 & -6.539 & 1.9 & 5\\
NGC 6242 & 165512.10-392902.2 & 253.800421 & -39.483953 & 7.7 & 3.0 & -6.790 & 1.8 & 7\\
NGC 6242 & 165512.12-393126.1 & 253.800519 & -39.523920 & 17.7 & 4.5 & -6.399 & 2.2 & 5\\
NGC 6242 & 165512.30-393114.6 & 253.801278 & -39.520725 & 3.5 & 2.5 & -7.126 & 1.2 & 10\\
NGC 6242 & 165512.33-392610.0 & 253.801389 & -39.436120 & 9.1 & 3.4 & -6.676 & 2.5 & 10\\
NGC 6242 & 165512.35-393218.5 & 253.801494 & -39.538476 & 6.2 & 3.2 & -6.879 & 3.7 & 10\\
NGC 6242 & 165512.38-393000.1 & 253.801617 & -39.500034 & 4.1 & 2.5 & -7.081 & 1.9 & 10\\
\enddata
\tablecomments{This table is available in its entirety (7,995 X-ray sources) in machine-readable form in the online journal. These {\it Chandra}-X-ray source positions and photometric quantities are provided by the {\it ACIS Extract} package. Column 1: Cluster name. Column 2: IAU designation. Columns 3-4: Right ascension and declination (in decimal degrees) for epoch J2000.0. The X-ray photometric quantities listed in Columns 5-8 are calculated in the $(0.5-8)$~keV band. Columns 5-6: Net counts and average of the upper and lower 1-$\sigma$ errors. Column 7: $\log$ of apparent photometric flux as the ratio of the net counts to the mean Auxiliary Response File value (product of the local effective area and quantum efficiency) and exposure time. Column 8: Background-corrected median photon energy. Column 9: Source class: Groups 5-10 (see Section~\ref{sec_membership}).}
\end{deluxetable*}

%\epsscale{1.15}
\begin{figure*}[ht!]
\plotone{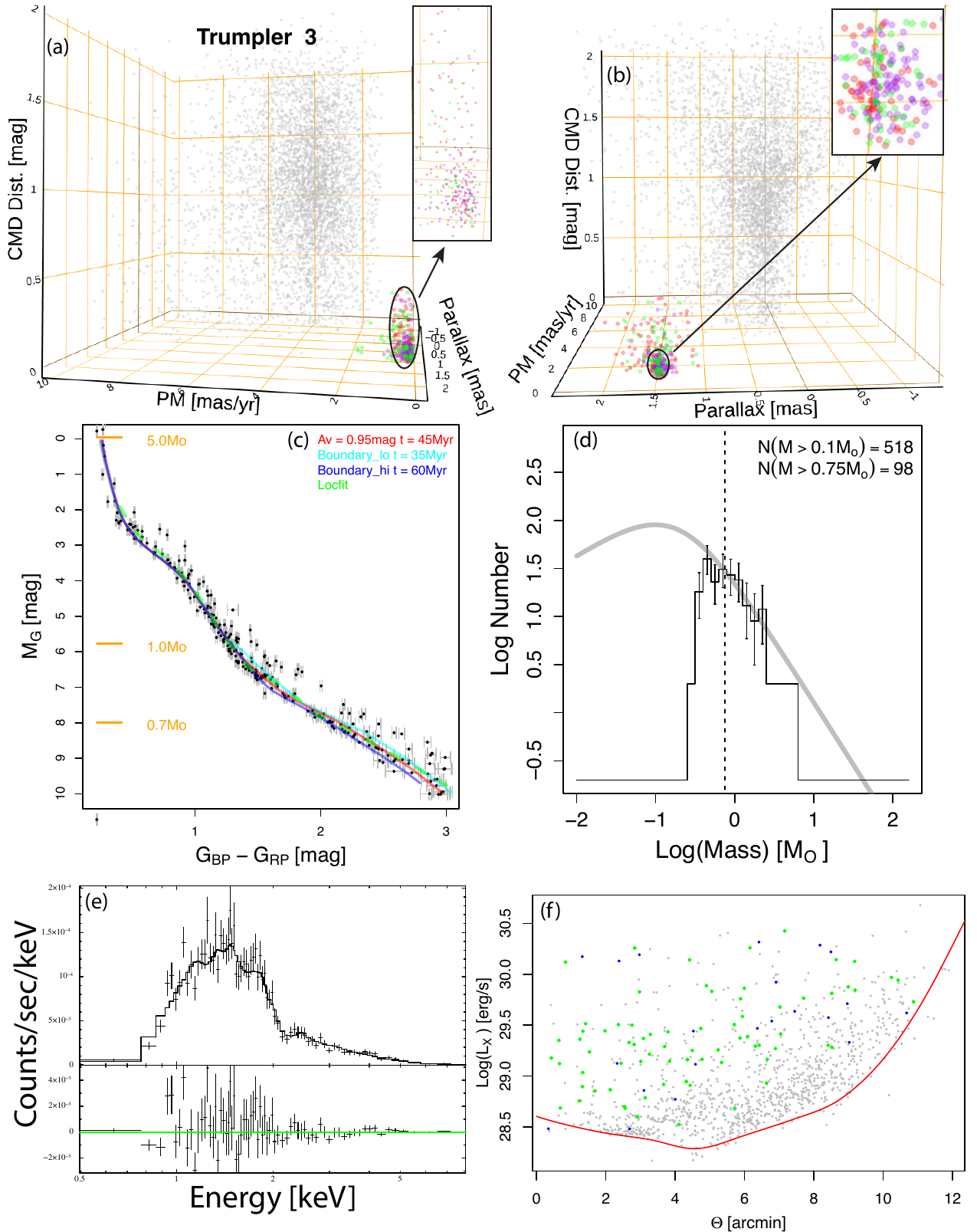}
\caption{Summary of cluster member identification and characterization for Trumpler 3. See Figure~\ref{fig:mem_NGC6242} for 
a detailed description. \label{fig:mem_TRUMPLER3}}
\end{figure*}

\begin{figure*}[ht!]
\plotone{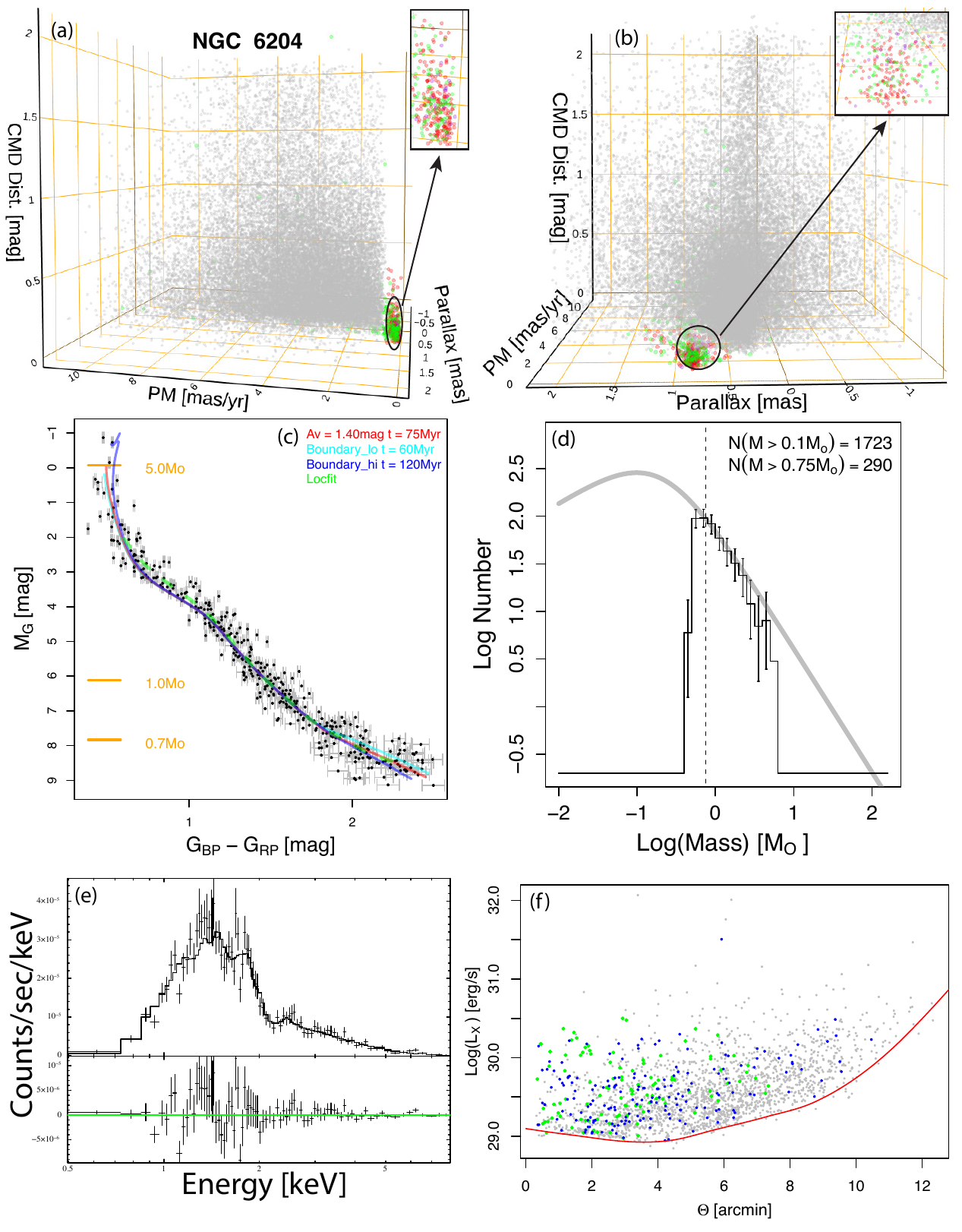}
\caption{Summary of cluster member identification and characterization for NGC 6204. See Figure~\ref{fig:mem_NGC6242} for 
a detailed description. \label{fig:mem_NGC6204}}
\end{figure*}

\begin{figure*}[ht!]
\plotone{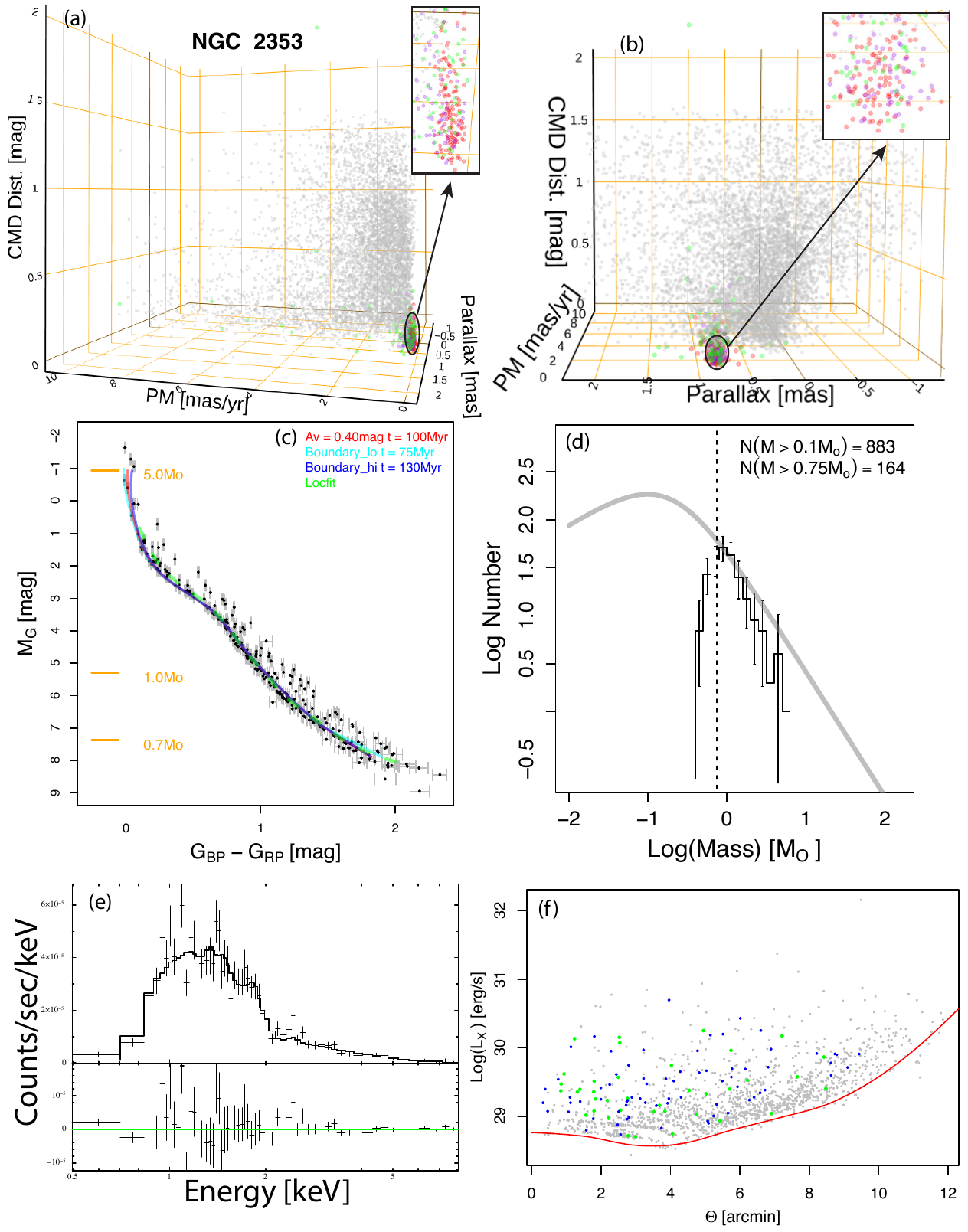}
\caption{Summary of cluster member identification and characterization for NGC 2353. See Figure~\ref{fig:mem_NGC6242} for 
a detailed description. \label{fig:mem_NGC2353}}
\end{figure*}

\begin{figure*}[ht!]
\plotone{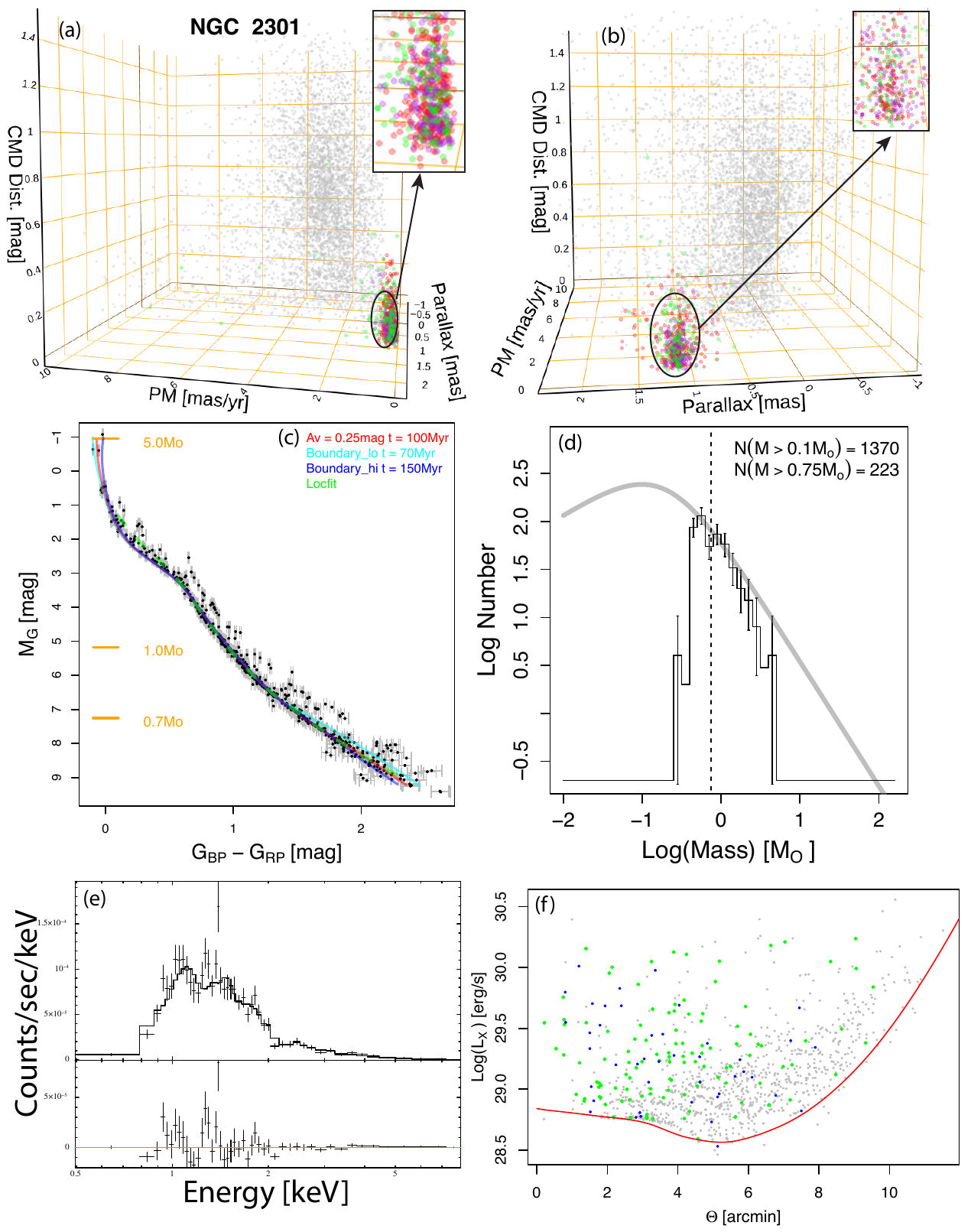}
\caption{Summary of cluster member identification and characterization for NGC 2301. See Figure~\ref{fig:mem_NGC6242} for 
a detailed description. \label{fig:mem_NGC2301}}
\end{figure*}

\begin{figure*}[ht!]
\plotone{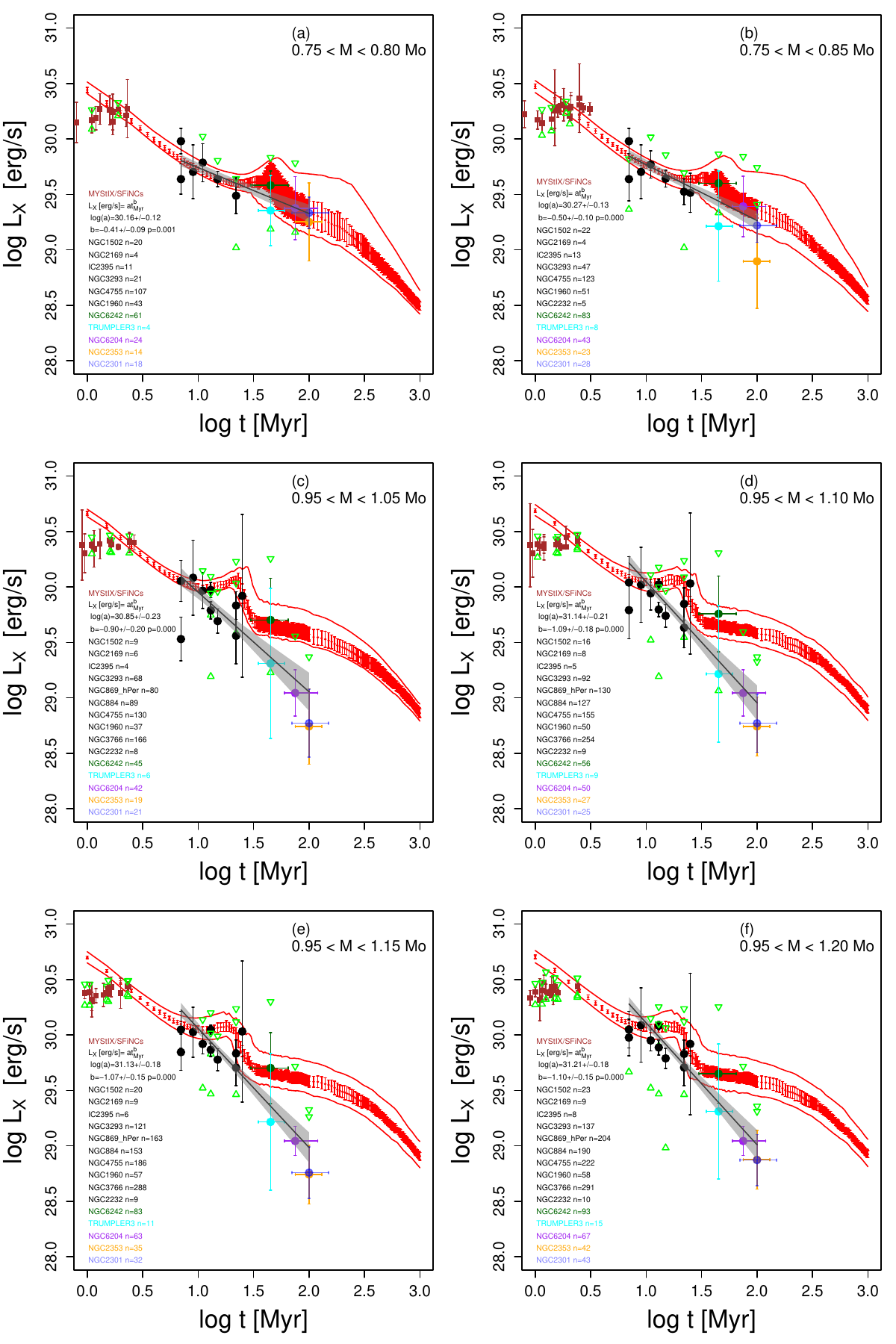}
\caption{Temporal evolution of X-ray luminosity across six stellar mass strata. See Figure~\ref{fig:lx_vs_t_main} for a detailed description. \label{fig:lx_vs_t_appendix}}
\end{figure*}

%% For this sample we use BibTeX plus aasjournalv7.bst to generate the
%% the bibliography. The sample7.bib file was populated from ADS. To
%% get the citations to show in the compiled file do the following:
%%
%% pdflatex sample7.tex
%% bibtext sample7
%% pdflatex sample7.tex
%% pdflatex sample7.tex

%%\bibliography{sample701}{}
\bibliography{my_bibliography}{}
\bibliographystyle{aasjournalv7}

%% This command is needed to show the entire author+affiliation list when
%% the collaboration and author truncation commands are used.  It has to
%% go at the end of the manuscript.
%\allauthors

%% Include this line if you are using the \added, \replaced, \deleted
%% commands to see a summary list of all changes at the end of the article.
%\listofchanges

\end{document}